# Practical and Efficient Verification of Entanglement with Incomplete Measurement Settings

Jiheon Seong,[1, *] Jin-Woo Kim,[1, 2, *] Seungchan Seo,[1] Seung-Hyun Nam,[3] Anindita Bera,[4]
Dariusz Chruściński,[5] June-Koo Kevin Rhee,[1] Heonoh Kim,[6, †] and Joonwoo Bae[1, ‡]

[1]*School of Electrical Engineering, Korea Advanced Institute of Science and Technology (KAIST),*
*291 Daehak-ro, Yuseong-gu, Daejeon 34141, Republic of Korea*
[2]*Photonic and Wireless Convergence Components Research Division,*
*Electronics and Telecommunications Research Institute (ETRI), Daejeon 34129, Republic of Korea*
[3]*Information & Electronics Research Institute, Korea Advanced Institute of Science and Technology, Daejeon, Republic of Korea*
[4]*Department of Mathematics, Birla Institute of Technology Mesra, Jharkhand 835215, India*
[5]*Institute of Physics, Faculty of Physics, Astronomy and Informatics,*
*Nicolaus Copernicus University, Grudziądzka 5/7, 87–100 Toruń, Poland*
[6]*Satellite Technology Research Center, Korea Advanced Institute of Science and Technology (KAIST),*
*291 Daehak-ro, Yuseong-gu, Daejeon 34141, Republic of Korea*

In this work, we present a practical and efficient framework for verifying entangled states when only a tomographically incomplete measurement setting is available—specifically, when access to observables is severely limited. We show how the experimental estimation of a small number of observables can be directly exploited to construct a large family of entanglement witnesses, enabling the efficient identification of entangled states. Moreover, we introduce an optimization approach, formulated as a semidefinite program, that systematically searches for those witnesses best suited to reveal entanglement under the given measurement constraints. We demonstrate the practicality of the approach in a proof-of-principle experiment with photon-polarization qubits, where entanglement is certified using only a fraction of the full measurement data. These results reveal the maximal usefulness of incomplete measurement settings for entanglement verification in realistic scenarios.

*Introduction.* Entanglement is a resource that enables quantum information processing, such as computing [1–3], communication [4–7], and metrology [8], and its verification is therefore of both fundamental and practical interest. The verification of entanglement may depend on whether the state of interest is identified and thus known in advance. When a state is known in advance after quantum state tomography, deciding whether it is entangled or separable is the well-defined problem of quantum state separability [9, 10]. Although the problem is computationally hard, known as NP-hard [11], theoretical tools have been developed, positive but not completely maps [12–14], the realignment criteria [15, 16], entropic separability [17], semidefinite programs, etc [18–20].

Entangled states can be experimentally verified even before quantum states are identified [21]. For example, an experimental violation of Bell inequalities concludes the presence of unknown entangled states [22]. In general, entanglement witnesses (EW) can detect entanglement of unknown states, and also can verify all entangled states [14, 23]. In experimental realizations, EWs can be realized as local observables, which then give a recipe for a measurement setting to detect entangled states, see e.g. [24]. In this way, experimental demonstrations of EWs have been presented, see e.g., [25–28].

Here, we approach the verification of entanglement in the opposite direction by asking how large the set of entangled states can be verified by a given measurement setting. Specifically, taking EWs as a tool for detecting entangled states, the question aims to characterize the set of EWs that can be realized by a single measurement setting; then, entangled states detected by those EWs can be verified by the measurement setting. One can expect that more measurement settings would verify a larger set of entangled states, see e.g., [29, 30]; a maximal measurement setting will realize quantum state tomography.

*Summary of results.* In this work, we present a practical, efficient method for verifying entangled states by constructing multiple EWs once a tomographically incomplete measurement setting is provided. Given measurement results that cannot reconstruct a state, we show how to verify entangled states by constructing multiple EWs from the results of that measurement setting. The construction includes both mirrored EWs and numerical optimization with measurement results, with the optimization formulated as a semidefinite program. It is also fairly straightforward to apply the results to multipartite and high-dimensional systems. We present a proof-of-principle demonstration of verifying two-qubit entangled states with polarization qubits.

We stress that our results are practically useful, as constraints on measurement settings often arise in realistic scenarios. Instances may include quantum communication between ground and moving stations, such as a satellite or an aircraft, where the simpler the measurement setup, the more practical it is; see, e.g., [31]. In fact, the full tomographic measurement setting may not be available. From a fundamental point of view, our results resolve the measurement settings for entanglement detection. The more measurements are applied, the larger the set of entangled states that can be verified; the construction of multiple EWs from a measurement setting is interpolated with quantum state tomography, applying a maximal setting for state identification.

* equally contributed
† quantumkho@kaist.ac.kr
‡ joonwoo.bae@kaist.ac.kr

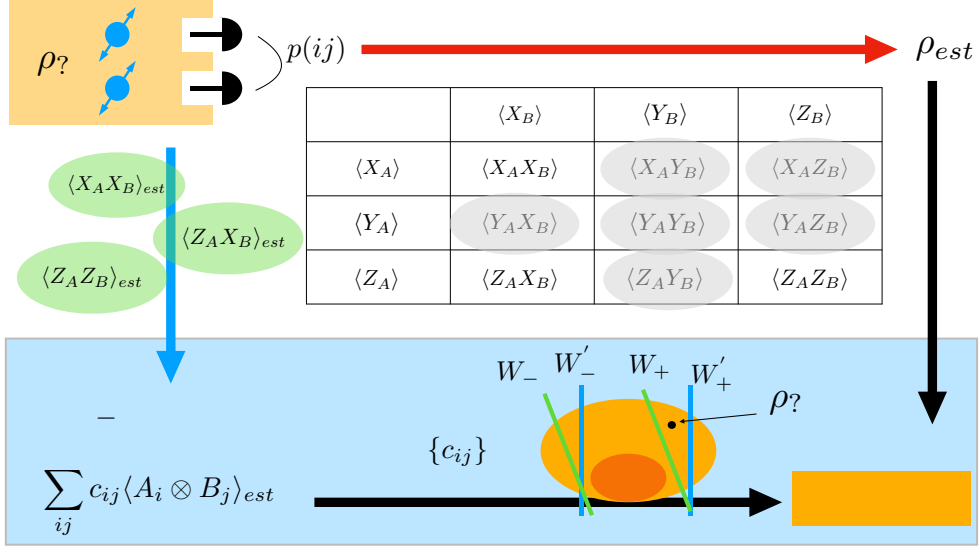

| | $\langle X_B \rangle$ | $\langle Y_B \rangle$ | $\langle Z_B \rangle$ |
|---|---|---|---|
| $\langle X_A \rangle$ | $\langle X_A X_B \rangle$ | $\langle X_A Y_B \rangle$ | $\langle X_A Z_B \rangle$ |
| $\langle Y_A \rangle$ | $\langle Y_A X_B \rangle$ | $\langle Y_A Y_B \rangle$ | $\langle Y_A Z_B \rangle$ |
| $\langle Z_A \rangle$ | $\langle Z_A X_B \rangle$ | $\langle Z_A Y_B \rangle$ | $\langle Z_A Z_B \rangle$ |

Figure 1. (Red-colored arrow) For an unknown state, denoted by $\rho_?$, measurement results of all correlators can characterize a state $\rho_{est}$ through state tomography. Theoretical tools such as the partial transpose can be used to determine if a state $\rho_{est}$ is entangled.
(Blue-colored box) In a realistic scenario such as long-distancce quantum communication, all experimental data may not be available (gray ellipsoid). Given a fraction of measurement results (gree ellipsoid) that cannot reconstruct a state, the present work (blue box) introduces a post-optimization protocol with incomplete measurement results. We collect experimental data, $\langle A_i \otimes B_j \rangle_{est}$, and consider their linear combination with coefficients $\{c_{ij}\}$, corresponding to an observable $S$ in Eq. (2). Each set $\{c_{ij}\}$ defines a pair of EWs, see Eq. (4); hence, numerous EWs are generated, including mirrored EWs, see Eq. (4). This work shows that a fraction of measurement results suffices to construct various EWs to verify the entanglement of an unknown state.

*Preliminaries.* Local measurements can be expressed in terms of correlators of canonical observables. For qubits, the Pauli observables $\mathbb{I}$, $X$, $Y$, and $Z$ form a basis, and thus estimating them provides tomographically complete measurement results, see Fig. 1. For bipartite systems, the collection of all correlators

$$\langle A \otimes B \rangle_{est} = \text{tr}[A \otimes B\rho] \ \text{ for } A, B \in \{\mathbb{I}, X, Y, Z\} \quad (1)$$

can reconstruct a two-qubit state $\rho$. As shown in Fig. 1, a measurement setting may generally be used to obtain a collection of correlators, which we exploit for the verification of entangled states.

*Results.* Throughout, we suppose that two-qubit ob-servables $A_i \otimes B_j$ for $A_i, B_j \in \{X, Y, Z\}$ are experimentally estimated and provided for an unknown state. Our goal is to construct EWs that may verify whether the state is entangled. To this end, let $S$ denote a linear combination of observables that are experimentally available and accessible.

$$S = \sum_{ij} c_{ij} A_i \otimes B_j \quad (2)$$

with real parameters $c_{ij}$, which are fully under manipulation to detect entangled states.

Let $C = [c_{ij}]$ denote a matrix with elements $c_{ij}$. We recall the operator norm of a matrix, $\|C\|_\infty = \max_k \lambda_k$, where $\{\lambda_k\}$ are singular values of a matrix $C$. For all separable states $\sigma_{\text{sep}}$, see Eq. (1), it holds that

$$-\|C\|_\infty \leq \sum_{ij} c_{ij} \langle A_i \otimes B_j \rangle_{est} \leq \|C\|_\infty. \quad (3)$$

We detail the proof in Appendix I.B. and see also related works, e.g., [32].

In fact, the inequalities above define two EWs:

$$W_\pm = \|C\|_\infty \mathbb{I} \pm S. \quad (4)$$

Note that they satisfy the relation $W_+ + W_- = 2\|C\|_\infty \mathbb{I}$ and hence define a pair of so-called mirrored EWs [33].

One can see that Eq. (4) provides a construction of the entire class of EWs parametrized by matrix elements $C = [c_{ij}]$. Hence, given that a fraction of experimentally accessible observables $A_i \otimes B_j$, we aim to search for a matrix $C$ to identify instances in which the inequalities in Eq. (3) are violated, or equivalently, to construct EWs detecting entangled states as in Eq. (4). For this purpose, we need to optimize a matrix $C$, see Fig. 1.

We now formulate the optimization problem on a matrix $C$ for verifying entangled states via the detection criteria in Eq. (3). We introduce a normalized expectation, denoted by NE: given a set of qubit observables, denoted by $\mathcal{O} = \{A_i \otimes B_j\}$, and their experimental estimations for an unknown state $\rho$, a normalized estimation parameter $\text{NE}_\rho[\mathcal{O}]$ is defined as,

$$\text{NE}_\rho[\mathcal{O}] \;=\; \max \left| \sum_{ij} c_{ij} \langle A_i \otimes B_j \rangle_{est} \right| \qquad (5)$$

$$\text{subject to} \qquad \|C\|_\infty = 1$$

where maximization runs over real parameters $\{c_{ij}\}$. It is clear that $\text{NE}_{\sigma_{\text{sep}}}[\mathcal{O}] \leq 1$ for all separable states $\sigma_{\text{sep}}$, and a state $\rho$ is entangled if $\text{NE}_\rho[\mathcal{O}] > 1$. The optimization problem in Eq. (5) can be solved numerically via the semidefinite program, which is detailed in Appendix I.C.

*Illustrations with two correlators.* Let us illustrate constructions of multiple EWs with an example of two correlators $XX$ and $ZZ$, where $XX := X \otimes X$ and $ZZ := Z \otimes Z$. In this case, we have $C = \text{diag}[\alpha, 0, \beta]$,

$$\|C\|_\infty = \max(|\alpha|, |\beta|).$$

Once the observables are estimated, it holds that for all separable states

$$-\|C\|_\infty \leq \alpha \langle XX \rangle_{est} + \beta \langle ZZ \rangle_{est} \leq \|C\|_\infty. \qquad (6)$$

Therefore, violations of either inequality for an unknown state imply that the state must be entangled.

Our framework for verifying entangled states works by searching for an optimal matrix $C$ in Eq. (6). The normalized estimation, from the inequalities above, is computed as follows,

$$\text{NE}_\rho[\mathcal{O}] = \max_{\alpha, \beta} \frac{|\alpha \langle XX \rangle_{est} + \beta \langle ZZ \rangle_{est}|}{\|C\|_\infty}, \qquad (7)$$

where $\mathcal{O} = \{XX, ZZ\}$ denotes a collection of observables. Note that for all separable states $\sigma_{\text{sep}}$ we have $\text{NE}_{\sigma_{\text{sep}}}[\mathcal{O}] \leq 1$, see Eq. (6). If the maximal value in Eq. (7) exceeds 1, the unknown state must be entangled. Interestingly, one can analytically solve the optimization,

$$\text{NE}_\rho[\mathcal{O}] = |\langle XX \rangle_{est}| + |\langle ZZ \rangle_{est}|.$$

Hence, we conclude that a two-qubit state must be entangled if it is found $|\langle XX \rangle_{est}| + |\langle ZZ \rangle_{est}| > 1$.

We remark that Eq. (7) obtained from the estimation of two observables can construct pairs of EWs,

$$W_\pm \;=\; \|C\|_\infty \mathbb{I} \pm (\alpha XX + \beta ZZ).$$

Various EWs can be obtained by choosing parameters $(\alpha, \beta)$, and are listed in Appendix II.A.

**Example.** Suppose a two-qubit entangled state in a chosen basis, $|\psi_\theta\rangle = \cos\theta \, |00\rangle + \sin\theta \, |11\rangle$, for $\theta \neq 0, \pi$, for which the estimation of two observables $XX$ and $ZZ$ is available. We may have experimental estimation: $\langle XX \rangle_{est} = \sin 2\theta$ and $\langle ZZ \rangle_{est} = 1$. Without loss of generality, we assume $|\beta| \geq |\alpha|$ so that the normalized estimation in Eq. (7)

$$\frac{|\alpha \langle XX \rangle_{est} + \beta \langle ZZ \rangle_{est}|}{|\beta|} = \left| 1 + \frac{\alpha}{\beta} \sin 2\theta \right|.$$

Taking $\alpha = \pm \beta$ one finds $1 + |\sin 2\theta| > 1$. We have shown that a two-qubit entangled state $|\psi_\theta\rangle$, while the basis is known, can be detected.

Note also that if local bases are not known in advance, two observables may not be sufficient to verify all entangled states. For instance, observables $XX$ and $ZZ$ vanish with respect to a maximally entangled state

$$|\varphi\rangle = \frac{1}{\sqrt{2}} (|0\rangle|+\rangle + |1\rangle|-\rangle)/\sqrt{2}$$

i.e.,

$$\text{tr}[XX|\varphi\rangle\langle\varphi|] = 0 \text{ and } \text{tr}[ZZ|\varphi\rangle\langle\varphi|] = 0,$$

where $|\pm\rangle = (|0\rangle \pm |1\rangle)/\sqrt{2}$. Clearly, two observables $XX$ and $ZZ$ do not suffice to determine if a state $|\varphi\rangle$ is entangled. An additional observable is required to verify the presence of entanglement, which will be demonstrated experimentally in what follows.

*Experimental Demonstrations.* We here consider a fraction of measurement results from two-photon experiments with polarization qubits, and demonstrate the verification of entanglement via the optimization in Eq. (5). Horizontal and vertical polarizations are denoted as the computational basis. We also demonstrate constructions of multiple EWs from tomographically incomplete measurements. In particular, we increase the number of accessible observables from two to four, see Table I.

The experimental setup is shown in Fig. 2. An entangled photon pair is generated in a quasi-phase matched PPKTP crystal designed for type-II phase matching, pumped by a single-mode laser with a center wavelength of $\lambda_p = 403.6$ nm. Entangled photon pairs are initially generated in a state $|\psi^-\rangle = (|01\rangle - |10\rangle)/\sqrt{2}$, which can be rotated to other Bell states by half-wave plates and quarter-wave plates. The photon pairs are separated into distinct spatial modes and detected with single-photon avalanche diodes. In each mode, the measurement basis can be independently chosen, and coincidence measurements are performed within a 2 ns window. This setup enables state manipulation and measurement for entanglement verification.

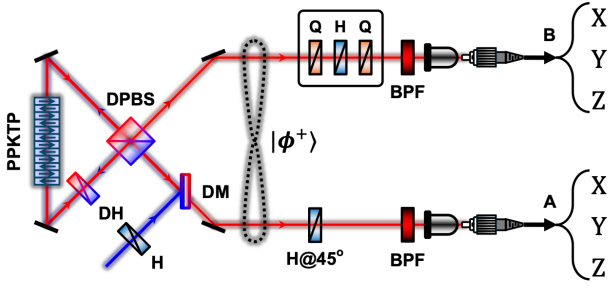

Figure 2. An experimental setup for generating two-photon entangled states and estimating correlators is shown, where optical elements are half-wave plates (H), quarter-wave plates (Q), dichroic mirrors (DM), dual-wavelength polarizing beam splitters (DPBS), dual-wavelength half-wave plates (DH), band-pass filters (BPF), and periodically poled KTiOPO4 (PPTKP). Entangled photon pairs are generated in a Sagnac interferometer, and then local operations are performed to prepare a state $|\phi^+\rangle$. A rotation of a polarization qubit can be realized with $Q$ and $H$, by which states $|\chi_1\rangle$ in Eq. (8) and $|\chi_3\rangle$ in Eq. (10) are prepared. Measurements are performed in bases $\{X, Y, Z\}$, from which correlators are estimated. The results of state tomography for states $|\chi_1\rangle$ and $|\chi_3\rangle$ are shown in Fig. 3.

In the experiment, we have prepared a state in the following, by varying an angle $\theta$,

$$|\chi_1\rangle = (I \otimes R_Y(\theta))|\phi^+\rangle, \quad R_Y(\theta) = \exp[-i\theta Y/2]. \quad (8)$$

For instance, for $\theta = \pi$, we have the state fidelity as $F = 98\%$, see Fig. 3. From the nine correlators, we consider three observables $XX$, $XY$, and $ZX$ as the accessible ones, and have estimation results experimentally,

$$\langle XX\rangle_{est} = -0.95, \quad \langle XY\rangle_{est} = 0.03, \quad \langle ZX\rangle_{est} = -0.96.$$

From these, we search for EWs for verifying entangled states by the optimization problem, see Eq. (5)

$$\max_{\alpha, \beta, \gamma} |\alpha\langle XX\rangle_{est} + \beta\langle XY\rangle_{est} + \gamma\langle ZX\rangle_{est}| \quad (9)$$

subject to $\left\| \begin{pmatrix} \alpha & \beta & 0 \\ 0 & 0 & 0 \\ \gamma & 0 & 0 \end{pmatrix} \right\|_\infty = 1.$

The optimization can be approached by the semidefinite program, and we obtain the optimal parameters: $\alpha = -0.70$, $\beta = 0.04$, and $\gamma = -0.71$, i.e.,

$$\mathbb{NE}_\rho[\mathcal{O}] = \max_{\alpha, \beta, \gamma} |-0.95\alpha + 0.03\beta - 0.96\gamma| = 1.35.$$

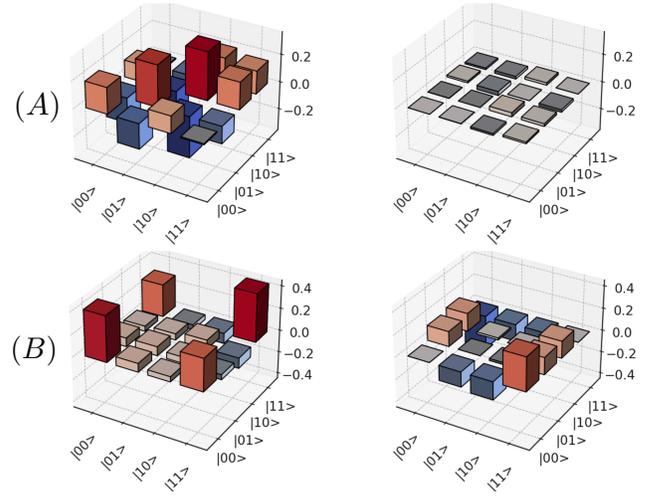

Figure 3. The state tomography is performed for polarization-entangled two-photon states: (A) a state $|\chi_1(\theta = \pi)\rangle$ in Eq. (8) and (B) a state $|\chi_3\rangle$ in Eq. (10). The left column shows real parts, and the right column shows imaginary parts. The fidelities are obtained as 98% and 95%, respectively.

Since $\mathbb{NE}_\rho[\mathcal{O}] > 1$, we unambiguously conclude an entangled state. Thus, we have shown that estimating three observables, $XX$, $XY$, and $ZX$, can verify an entangled state in Eq. (8) for $\theta = \pi$.

We have also considered another instance, a maximally entangled state,

$$|\chi_3\rangle = (I \otimes V)|\phi^+\rangle, \quad V = \frac{1}{\sqrt{2}}(\mathbb{I} + i(\cos\theta X + \sin\theta Z)) \quad (10)$$

with $\theta = 7/9\,\pi$. An experimental realization of the state has a state fidelity $F = 95\%$, see Fig. 3. From two to four observables, we compute Eq. (5) and find that more measurement settings yield larger violations. The results are summarized in Table I. Two observables suffice to verify an entangled state, and a larger violation is shown with more observables. A larger violation implies that entanglement can be verified even when the state contains a higher noise rate.

*Generalizations.* The framework of detecting entangled states with limited correlators, see Eq. (5), can be generalized to high-dimensional quantum systems. In this case, local observables $A_i$ and $B_j$ may be chosen as $d$-dimensional Hermitian operators, e.g., generators of $SU(d)$ [34]: $G_i^A$ ($i = 1, \ldots, d_A^2$) and $G_j^B$ ($j = 1, \ldots, d_B^2$) with $Tr(G_k^A G_\ell^A) = d_A \delta_{kl}$ and the similarly for $G_j^B$. Now, for any $d_A \otimes d_B$ separable state one has [32]

$$\left| \sum_{ij} c_{ij}\langle G_i^A G_j^B\rangle_{est} \right| \leq \sqrt{(d_A - 1)(d_B - 1)} \|C\|_\infty, \quad (11)$$

which reduces to (3) for $d_A = d_B = 2$.

| Normalized estimation | Max. |
|---|---|
| $1.0\langle XX\rangle_{est} + 1.0\langle ZZ\rangle_{est}$ | 1.36 |
| $1.0\langle XX\rangle_{est} + 0.67\langle ZY\rangle_{est} + 0.74\langle ZZ\rangle_{est}$ | 1.60 |
| $0.76\langle XX\rangle_{est} - 0.65\langle YX\rangle_{est} + 0.67\langle ZY\rangle_{est} + 0.74\langle ZZ\rangle_{est}$ | 1.82 |

Table I. Expectation values of correlators from a state $|\chi_3\rangle$, see Fig. 3 (B), are obtained in the proof-of-principle demonstration. Given correlators above, normalized estimations in Eq. (5) are computed. In all cases, the maximum is larger than 1, and more observables can find larger violations; hence, more observables can detect a larger set of entangled states.

Finally, the detection method can be generalized to multipartite and high-dimensional systems. For multipartite qubit states, we formulate Eq. (5) for observables of multipartite systems,

$$\mathbb{NE}_\rho[\mathcal{O}] = \max_{c_{i_1\cdots i_n}} \frac{1}{\lambda_{\max}} \left| \sum_{i_1\cdots i_n} c_{i_1\cdots i_n} \langle o_{i_1}^{[1]} \otimes \cdots \otimes o_{i_n}^{[n]} \rangle_{est} \right| \quad (12)$$

where $o_{i_k}^{[k]}$ denotes a local observable of the $k$-th system for $k = 1, \cdots, n$, and

$$\lambda_{\max} = \max_{|\psi_{\text{prod}}\rangle} \langle \psi_{\text{prod}} | \sum c_{i_1\cdots i_n} o_{i_1}^{[1]} \otimes \cdots \otimes o_{i_n}^{[n]} | \psi_{\text{prod}} \rangle \quad (13)$$

where $|\psi_{\text{prod}}\rangle = |e_1 \cdots e_n\rangle$, i.e., the maximal expectation of an observable of a multipartite system over product states. Note also that the construction with EWs can be generalized to correlation tensors in multipartite systems [35]. The maximization in Eq. (13) can be computed numerically [36]. Entangled states are detected when Eq. (12) is larger than 1. Since $\lambda_{\max}$ in Eq. (13) is computed for fully separable states, the method in Eq. (12) can detect entangled states that are not fully separable. For multipartite systems, mirrored EWs are efficient as there are mirrored pairs of optimal EWs [37].

We can exploit constraints on a state $|\psi\rangle$ in the computation of $\lambda_{\max}$ in Eq. (13) to detect various structures of entangled states. For instance, an $n$-partite state in cuts $A|B|C$,

$$|\psi\rangle_{1\cdots n} = |a\rangle_A |b\rangle_B |c\rangle_C$$

where $A$, $B$, and $C$ are disjoint subjects with indices $\{1, \cdots, n\}$, can be used to detect those entangled states that are not separable in the cuts, called 3-separable states. In general, entangled states that are not $k$-separable can be detected by computing $\lambda_{\max}$ with $k$-separable states. The proposed method can be applied to detecting entangled states that are not $k$-separable.

*Conclusion.* We have presented a practical, efficient method for verifying entangled states using tomographically incomplete measurements. Our results here do not demonstrate EWs, but we present a framework for constructing multiple EWs to verify a maximal set of entangled states, provided limited access to the correlation data, i.e., a measurement setting that yields tomographically incomplete results. The construction of EW also exploits mirrored EWs, which have both upper and lower bounds satisfied by all separable states. Hence, our results provide a new, efficient, and practical approach to verifying entangled states in realistic scenarios. We envisage that the proposed method applies to various realistic scenarios for distributing entangled states, particularly when the measurement setting is limited and thus tomographically incomplete, e.g., at ground and moving stations that aim to distribute entanglement, where state tomography is not practically feasible.

### ACKNOWLEDGEMENT

J.S., A.B., D.C., and J.B. thank Beatrix C. Hiesmayr for useful discussions.

J.S., S.S., S.N., and J.B. are supported by the National Research Foundation of Korea (Grant No. NRF-2021R1A2C2006309, NRF-2022M1A3C2069728), the Institute for Information & Communication Technology Promotion (IITP) (RS-2023-00229524, RS-2025-02304540, RS-2025-25464876).

J.S., S.S., J.B., J.K., J.K.R., and H.K. are supported by the KAIST Quantum+X Convergence R&D Project.

A.B. acknowledges the support received for this research from the research grant sanctioned by the National Board for Higher Mathematics (NBHM), Department of Atomic Energy (DAE), Government of India, under sanction letter no: 02011/32/2025/NBHM(R.P.)/R&D II/9677.

D.C. was supported by the Polish National Science Centre project No. 2024/55/B/ST2/01781.

# Supplemental Material: Experimental Detection of Entanglement with Tomographically Incomplete Measurements

## CONTENTS

# I. OBSERVABLES FOR DETECTING ENTANGLED STATES

## A. Definitions

Consider a bipartite quantum system $\mathcal{H}_A \otimes \mathcal{H}_B$ where $\mathcal{H}_A \cong \mathbb{C}^d$ and $\mathcal{H}_B \cong \mathbb{C}^d$ for some dimension $d \geq 2$. We are interested in detecting entanglement in a bipartite state $\rho$ acting on $\mathcal{H}_A \otimes \mathcal{H}_B$ using only local measurements.

**Definition 1** (Entanglement witness (EW)). An EW is a Hermitian operator $W$ such that:

(i) $\mathrm{tr}[W\sigma_{\mathrm{sep}}] \geq 0$ for all separable states $\sigma_{\mathrm{sep}}$, and

(ii) there are some entangled states $\rho$ such that $\mathrm{tr}[W\rho] < 0$.

Let $\mathcal{O} = \{A_i \otimes B_j\}$ be a set of local observables, where $A_i$ acts on $\mathcal{H}_A$ and $B_j$ acts on $\mathcal{H}_B$. Each observable $A_i$ and $B_j$ is Hermitian, i.e., $A_i = A_i^\dagger$ and $B_j = B_j^\dagger$. In the following, we consider an observable

$$S = \sum_{i,j} c_{ij} A_i \otimes B_j \tag{1}$$

where $\{c_{ij}\}$ are real.

## B. Separable bounds of observables

### 1. Bipartite case

**Definition 2** (Operator norm). For a matrix $C = [c_{ij}]$, its operator norm is defined as

$$\|C\|_\infty = \sqrt{\max \text{ eigenvalue of } (C^\dagger C)}.$$

An operator in Eq. (1) can be rewritten in terms of orthogonal matrices,

$$S = \sum_{i=1}^{d^2-1} \sum_{j=1}^{d^2-1} c_{ij} G_i^A \otimes G_j^B, \tag{2}$$

where $\{G_i^A\}_{i=0}^{d^2-1}$ ($\{G_i^B\}_{i=0}^{d^2-1}$) is an orthogonal operator basis on each local Hilbert space $\mathcal{H}_A$ ($\mathcal{H}_B$), normalized as

$$\mathrm{tr}\left(G_k^{A\dagger} G_l^A\right) = d\,\delta_{kl}, \tag{3}$$

with the identity singled out as

$$G_0^A = \mathbb{I}, \quad \text{and} \quad \mathrm{tr}\left(G_i^A\right) = 0 \text{ for } i = 1, \cdots, d-1, \tag{4}$$

and similarly for $G_j^B$. We again denote by $C = [c_{ij}]$ the $(d^2-1) \times (d^2-1)$ coefficient matrix appearing in the correlator expansion of $S$. Then we prove the following proposition for the general qudit-qudit bipartite case using an orthogonal operator basis decomposition, see also Refs. [1–3].

**Proposition 1** (Separable bounds). For all separable states $\sigma_{\mathrm{sep}}$ and $S$ in Eq. (2), it holds that

$$-(d-1)\|C\|_\infty \leq \mathrm{tr}[S\,\sigma_{\mathrm{sep}}] \leq (d-1)\|C\|_\infty. \tag{5}$$

*Proof.* Let

$$\sigma_{\mathrm{sep}} = \sum_k p_k \rho_k^A \otimes \rho_k^B. \tag{6}$$

Then

$$\text{tr}\left[S\sigma_{\text{sep}}\right] = \sum_k p_k \sum_{i,j=1}^{d^2-1} c_{ij}\ \text{tr}\left(\rho_k^A G_i^A\right)\ \text{tr}\left(\rho_k^B G_j^B\right). \tag{7}$$

Define real vectors

$$(\vec{v}_k)_i = \text{tr}\left(\rho_k^A G_i^A\right), \qquad (\vec{w}_k)_j = \text{tr}\left(\rho_k^B G_j^B\right), \qquad i,j = 1,\ldots,d^2-1. \tag{8}$$

Then

$$\text{tr}\left[S\sigma_{\text{sep}}\right] = \sum_k p_k\ \vec{v}_k^{\mathsf{T}} C\ \vec{w}_k. \tag{9}$$

Since $\{G_i^A\}_{i=0}^{d^2-1}$ is an orthogonal operator basis with $\text{tr}(G_k^{A\dagger} G_l^A) = d\delta_{kl}$, any state admits the expansion

$$\rho = \frac{1}{d}\sum_{i=0}^{d^2-1} \text{tr}\left(\rho G_i^A\right) G_i^A, \tag{10}$$

and orthogonality yields the following identity

$$\text{tr}\left(\rho^2\right) = \frac{1}{d}\sum_{i=0}^{d^2-1} |\text{tr}\left(\rho G_i^A\right)|^2. \tag{11}$$

Now single out the identity component by choosing $G_0^A = \mathbb{I}$. For any pure state $\rho$, we have $\text{tr}(\rho^2) = 1$ and $\text{tr}(\rho G_0^A) = \text{tr}(\rho\mathbb{I}) = 1$, so

$$\sum_{i=1}^{d^2-1} |\text{tr}\left(\rho G_i^A\right)|^2 = d\ \text{tr}\left(\rho^2\right) - |\text{tr}\left(\rho G_0^A\right)|^2 = d-1, \tag{12}$$

and similarly for $\{G_i^B\}_{i=0}^{d^2-1}$. Therefore every $\vec{v}_k$ and $\vec{w}_k$ consisting of the traceless components satisfies

$$\|\vec{v}_k\|_2 \leq \sqrt{d-1}, \qquad \|\vec{w}_k\|_2 \leq \sqrt{d-1}. \tag{13}$$

Using the Cauchy–Schwarz and $L_2$-norm norm inequalities,

$$|\vec{v}_k^{\mathsf{T}} C \vec{w}_k| \leq \|C\|_\infty \|\vec{v}_k\|_2 \|\vec{w}_k\|_2 \leq (d-1)\|C\|_\infty. \tag{14}$$

Finally,

$$|\text{tr}\left[S\sigma_{\text{sep}}\right]| \leq \sum_k p_k |\vec{v}_k^{\mathsf{T}} C \vec{w}_k| \leq (d-1)\|C\|_\infty, \tag{15}$$

which proves the desired separable bound. ∎

**Corollary 1** (Two correlators). *For two correlators $XX$ and $ZZ$ with real parameters $\alpha$ and $\beta$, for all separable states we have*

$$-\max(|\alpha|,|\beta|) \leq \alpha\langle XX\rangle + \beta\langle ZZ\rangle \leq \max(|\alpha|,|\beta|). \tag{16}$$

*Violations of either inequality imply that the state must be entangled.*

*Proof.* For $S = \alpha(X \otimes X) + \beta(Z \otimes Z)$, the coefficient matrix is

$$C = \begin{pmatrix} \alpha & 0 \\ 0 & \beta \end{pmatrix},$$

and $\|C\|_\infty = \max(|\alpha|,|\beta|)$. The result follows directly from Proposition 1 (equation (5)). ∎

## 2. Multipartite case

For an $n$-partite system $\mathcal{H}_1 \otimes \cdots \otimes \mathcal{H}_n$, we extend the normalized estimation to multipartite observables. Let $o_{i_k}^{[k]}$ denote a local observable of the $k$-th system for $k = 1, \cdots, n$, and consider the set of local observables $\mathcal{O} = \{o_{i_1}^{[1]} \otimes \cdots \otimes o_{i_n}^{[n]}\}$.

The normalized estimation for a multipartite state $\rho$ is defined as

$$\mathbb{NE}_\rho[\mathcal{O}] = \max_{c_{i_1 \cdots i_n}} \frac{1}{\lambda_{\max}} \left| \sum_{i_1 \cdots i_n} c_{i_1 \cdots i_n} \operatorname{tr} \left[ o_{i_1}^{[1]} \otimes \cdots \otimes o_{i_n}^{[n]} \rho \right] \right|, \tag{17}$$

where $c_{i_1 \cdots i_n} \in \mathbb{R}$ are real coefficients, and

$$\lambda_{\max} = \max_{|\psi_{\text{prod}}\rangle} \left\langle \psi_{\text{prod}} \left| \sum_{i_1 \cdots i_n} c_{i_1 \cdots i_n} o_{i_1}^{[1]} \otimes \cdots \otimes o_{i_n}^{[n]} \right| \psi_{\text{prod}} \right\rangle, \tag{18}$$

where the maximization runs over all product states $|\psi_{\text{prod}}\rangle = |\psi_1\rangle \otimes \cdots \otimes |\psi_n\rangle$.

The computation of $\lambda_{\max}$ in the multipartite case can be efficiently performed using the Separability Power Iteration (SPI) method [4]. The SPI algorithm iteratively optimizes over each subsystem, converging to the maximal separability eigenvalue. The key idea is based on the cascaded structure of separable states, where the optimization over an $n$-partite operator can be reduced to an optimization over an $(n-1)$-partite reduced operator, as shown in the forward iteration theorem. The algorithm guarantees convergence to at least a local maximum, and by using multiple starting vectors from an operator basis, global convergence to the maximal separability eigenvalue is ensured.

### C. SDP formulation

Before turning to explicit examples, we first formalize the quantity $\mathbb{NE}_\rho(\mathcal{O})$ and introduce its equivalent SDP formulation. This provides a unified framework for understanding how subsets of Pauli measurements give rise to EWs which can be efficiently found via SDP. We first formally state that $\mathbb{NE}_\rho[\mathcal{O}] > 1$ leads to the conclusion that $\rho$ is entangled:

---

**Proposition 2** (Normalized estimation). *For a bipartite state $\rho$ in $\mathcal{H}_A \otimes \mathcal{H}_B$, with $\dim \mathcal{H}_A = \dim \mathcal{H}_B = d$, a normalized estimation for a set of local observables $\mathcal{O}$ which satisfies Eq. (3) is defined as,*

$$\mathbb{NE}_\rho[\mathcal{O}] = \max \left| \sum_{ij} c_{ij} \langle A_i \otimes B_j \rangle_{\text{est}} \right| \tag{19}$$

$$\text{subject to} \qquad \|C\|_\infty = \frac{1}{d-1}.$$

*Then $\rho$ is entangled if $\mathbb{NE}_\rho[\mathcal{O}] > 1$.*

*Proof.* We can construct two EWs from the separable bound on $S = \sum_{ij} c_{ij} A_i \otimes B_j$ in Proposition 1, namely $W_+$ and $W_-$ given by the following formulae:

$$W_+ = (d-1)\|C\|_\infty \mathbb{I} + S \quad \text{and} \quad W_- = (d-1)\|C\|_\infty \mathbb{I} - S. \tag{20}$$

Then $\mathbb{NE}_\rho[\mathcal{O}] > 1$ if and only if

$$\operatorname{tr}[W_+\rho] < 0 \quad \text{or} \quad \operatorname{tr}[W_-\rho] < 0, \tag{21}$$

thus implying that $\rho$ is detected by $W_+$ or $W_-$, which means that $\rho$ is entangled. ■

**Proposition 3** (SDP formulation of $\mathbb{NE}_\rho(\mathcal{O})$). *Let $\mathcal{O} = \{A_i \otimes B_j\}$ be a set of local Pauli measurements which satisfies Eq. (3) and let $\langle A_i \otimes B_j \rangle_{\text{est}}$ denote the estimated expectation values. Consider the following SDP problem*

$$\max \quad \left| \sum_{ij} c_{ij} \langle A_i \otimes B_j \rangle_{\text{est}} \right|, \tag{22}$$

$$\text{subject to} \quad \begin{pmatrix} \sqrt{d-1}\, \mathbb{I} & C \\ C^\top & \sqrt{d-1}\, \mathbb{I} \end{pmatrix} \succeq 0.$$

*The maximum above is equal to Eq. (25).*

*Proof.* Note that two constraints in the following are equivalent:

$$(d-1)\|C\|_\infty \leq 1 \quad \Longleftrightarrow \quad \begin{pmatrix} \sqrt{d-1}\, \mathbb{I} & C \\ C^\top & \sqrt{d-1}\, \mathbb{I} \end{pmatrix} \succeq 0. \tag{23}$$

Since the optimizer lies on the boundary, the maximization occurs when $\|C\|_\infty = 1/(d-1)$. ∎

### D. Closed-form normalized estimation for a few observables

In this subsection we focus on measurement sets $\mathcal{O}$ consisting of a small number of Pauli products. In these low-cardinality cases, the geometric arrangement of the selected measurements inside the $3 \times 3$ Pauli grid (see Fig. 1 for two-measurement patterns and Fig. 2 for three-measurement patterns) completely determines the structure of the optimization problem

$$\mathbb{NE}_\rho[\mathcal{O}] \;=\; \max \left| \sum_{ij} c_{ij} \langle A_i \otimes B_j \rangle_{\text{est}} \right| \tag{24}$$

$$\text{subject to} \quad \|C\|_\infty = 1.$$

where $C = [c_{ij}]$ is the coefficient matrix of $S$ in the fixed local Pauli basis. Collections of a few observables are shown in Figs. 1 and 2.

(a) Two correlators in a line          (b) Two diagonal correlators          (c) Two unadjacent correlators

Figure 1: Examples of two correlators out of $\{X, Y, Z\} \otimes \{X, Y, Z\}$. Note that two cases in (b) and (c) are local unitarily equivalent.

(a) Three correlators in a line  (b) Three diagonal correlators  (c) L-shape correlators  (d) Three unadjacent correlators

Figure 2: Four representative 3-measurement patterns on the Pauli-product grid $\{X, Y, Z\} \otimes \{X, Y, Z\}$.

### 1. Two Measurements

1. **Two correlators in a line, see Fig. 1 (a).**

   Note that for a measurement set $\mathcal{O} = \{XX, XY\}$

   $$S = \alpha\, XX + \beta\, XY, = X \otimes (\alpha X + \beta Y),\tag{25}$$

   so $S$ is merely a local unitary rotation of the single observable $XX$, which implies that we obtain

   $$|\langle S\rangle_\rho| \leq 1,\tag{26}$$

   for every state $\rho$ and the optimization condition $\|C\|_\infty = \sqrt{\alpha^2 + \beta^2} = 1$. Therefore the normalized estimation is always

   $$\mathbb{NE}_\rho[\{XX, XY\}] \leq 1,\tag{27}$$

   which shows this measurement set cannot detect entanglement.

2. **Two diagonal correlators, see Fig. 1 (b).**

   For any observable from a measurement set $\mathcal{O} = \{XX, ZZ\}$

   $$S = \alpha\, XX + \beta\, YY,\tag{28}$$

   the coefficient matrix is

   $$C = \begin{pmatrix} \alpha & 0 & 0 \\ 0 & \beta & 0 \\ 0 & 0 & 0 \end{pmatrix}, \qquad \|C\|_\infty = \max\{|\alpha|, |\beta|\}.\tag{29}$$

   Thus the normalized estimation with condition $\|C\|_\infty = 1$ is

   $$\mathbb{NE}_\rho[\{XX, ZZ\}] = \max |\alpha\, \langle XX\rangle_\rho + \beta\, \langle ZZ\rangle_\rho|.\tag{30}$$

   When $|\alpha| = |\beta| = 1$, we attain the following maximum value

   $$\mathbb{NE}_\rho[\{XX, ZZ\}] = |\langle XX\rangle_\rho| + |\langle ZZ\rangle_\rho|.\tag{31}$$

   Hence any value $> 1$ certifies entanglement. Note that all measurement sets which do not share a local Pauli operator are locally equivalent to measurement set $\mathcal{O} = \{XX, ZZ\}$.

### 2. Three Measurements

1. **Three correlators in a line, see Fig. 2 (a).**

   Note that for a measurement set $\mathcal{O} = \{XX, XY, XZ\}$

   $$S = \alpha\, XX + \beta\, XY + \gamma\, XZ = X \otimes (\alpha X + \beta Y + \gamma Z).\tag{32}$$

so $S$ is just a local unitary rotation of the single observable $XX$. Since

$$|\langle S\rangle_\rho| \leq 1 \tag{33}$$

for every state $\rho$ and the optimization condition $\|C\|_\infty = \sqrt{\alpha^2 + \beta^2 + \gamma^2} = 1$, the normalized estimation satisfies

$$\mathbb{NE}_\rho[\{XX, XY, XZ\}] \leq 1, \tag{34}$$

which implies that this measurement set cannot detect entanglement.

2. **Three diagonal correlators, see Fig. 2 (b).**

Consider an observable from a measurement set $\mathcal{O} = \{XX, YY, ZZ\}$

$$S = \alpha\, XX + \beta\, YY + \gamma\, ZZ. \tag{35}$$

The coefficient matrix is diagonal:

$$C = \begin{pmatrix} \alpha & 0 & 0 \\ 0 & \beta & 0 \\ 0 & 0 & \gamma \end{pmatrix}, \qquad \|C\|_\infty = \max\{|\alpha|, |\beta|, |\gamma|\}. \tag{36}$$

Thus the normalized estimation with condition $\|C\|_\infty = 1$ is

$$\mathbb{NE}_\rho[\mathcal{O}] = \max |\alpha\langle XX\rangle_\rho + \beta\langle YY\rangle_\rho + \gamma\langle ZZ\rangle_\rho|. \tag{37}$$

The optimum is achieved by $|\alpha| = |\beta| = |\gamma| = 1$. Therefore

$$\mathbb{NE}_\rho[\{XX, YY, ZZ\}] = |\langle XX\rangle_\rho| + |\langle YY\rangle_\rho| + |\langle ZZ\rangle_\rho|. \tag{38}$$

For the Bell state, all three correlators are $\pm 1$, giving

$$|\langle XX\rangle| + |\langle YY\rangle| + |\langle ZZ\rangle| = 3. \tag{39}$$

Hence any value $> 1$ certifies entanglement. The table I below shows all 3-measurement sets locally equivalent to $\{XX, YY, ZZ\}$:

| # | Measurement set $\mathcal{O}$ | $U_A \otimes U_B$ |
|---|---|---|
| 1 | $\{XX, YY, ZZ\}$ | $I \otimes I$ |
| 2 | $\{XX, YZ, ZY\}$ | $I \otimes (HSH)$ |
| 3 | $\{YY, XZ, ZX\}$ | $I \otimes H$ |
| 4 | $\{ZZ, XY, YX\}$ | $I \otimes S$ |
| 5 | $\{XY, YZ, ZX\}$ | $I \otimes (HS)$ |
| 6 | $\{XZ, YX, ZY\}$ | $I \otimes (SH)$ |

Table I: All 3-measurement 3-cycle sets locally equivalent to the representative set $\{XX, YY, ZZ\}$, along with corresponding local Cliffords $U_A \otimes U_B$.

3. **L-shape correlators, see Fig. 2 (c).**

Consider the measurement set $\mathcal{O} = \{XX,\ XZ,\ ZX\}$ and an observable from $\mathcal{O}$

$$S = \alpha\, XX + \beta\, XZ + \gamma\, ZX. \tag{40}$$

A calculation shows that the operator norm of $C$ is

$$\|C\|_\infty = \sqrt{\lambda_+}, \qquad \lambda_+ = \frac{T + \sqrt{T^2 - 4\beta^2\gamma^2}}{2}, \tag{41}$$

where $T = \alpha^2 + \beta^2 + \gamma^2$. Therefore the normalized estimation for this L-shape pattern reduces to the exact optimization problem

$$\mathrm{N}\mathbb{E}_\rho[\{XX, XZ, ZX\}] = \max |a\alpha + b\beta + c\gamma|, \tag{42}$$

with condition on $\alpha, \beta$ and $\gamma$ that

$$\beta^2\gamma^2 = (\alpha^2 + \beta^2 + \gamma^2) - 1, \tag{43}$$

which is equivalent to $\|C\|_\infty = 1$.

| # | Measurement set $\mathcal{O}$ | $U_A \otimes U_B$ | # | Measurement set $\mathcal{O}$ | $U_A \otimes U_B$ |
|---|---|---|---|---|---|
| 1 | $\{XX, XZ, ZX\}$ | $I \otimes I$ | 19 | $\{YX, YZ, XX\}$ | $(SH) \otimes I$ |
| 2 | $\{XX, XY, ZX\}$ | $I \otimes (HSH)$ | 20 | $\{YX, YY, XX\}$ | $(SH) \otimes (HSH)$ |
| 3 | $\{XY, XZ, ZY\}$ | $I \otimes S$ | 21 | $\{YY, YZ, YX\}$ | $(SH) \otimes S$ |
| 4 | $\{XX, XY, ZY\}$ | $I \otimes (SH)$ | 22 | $\{YX, YY, YX\}$ | $(SH) \otimes (SH)$ |
| 5 | $\{XY, XZ, ZZ\}$ | $I \otimes (HS)$ | 23 | $\{YY, YZ, ZX\}$ | $(SH) \otimes (HS)$ |
| 6 | $\{XX, XZ, ZZ\}$ | $I \otimes H$ | 24 | $\{YX, YZ, ZX\}$ | $(SH) \otimes H$ |
| 7 | $\{ZX, ZZ, XX\}$ | $H \otimes I$ | 25 | $\{ZX, ZZ, YX\}$ | $(HS) \otimes I$ |
| 8 | $\{ZX, ZY, XX\}$ | $H \otimes (HSH)$ | 26 | $\{ZX, ZY, YX\}$ | $(HS) \otimes (HSH)$ |
| 9 | $\{ZY, ZZ, YX\}$ | $H \otimes S$ | 27 | $\{ZY, ZZ, YY\}$ | $(HS) \otimes S$ |
| 10 | $\{ZX, ZY, YX\}$ | $H \otimes (SH)$ | 28 | $\{ZX, ZY, YY\}$ | $(HS) \otimes (SH)$ |
| 11 | $\{ZY, ZZ, ZX\}$ | $H \otimes (HS)$ | 29 | $\{ZY, ZZ, YX\}$ | $(HS) \otimes (HS)$ |
| 12 | $\{ZX, ZZ, ZX\}$ | $H \otimes H$ | 30 | $\{ZX, ZZ, YX\}$ | $(HS) \otimes H$ |
| 13 | $\{YX, YZ, ZX\}$ | $S \otimes I$ | 31 | $\{XX, XZ, YX\}$ | $(HSH) \otimes I$ |
| 14 | $\{YX, YY, ZX\}$ | $S \otimes (HSH)$ | 32 | $\{XX, XY, YX\}$ | $(HSH) \otimes (HSH)$ |
| 15 | $\{YY, YZ, ZY\}$ | $S \otimes S$ | 33 | $\{XY, XZ, ZX\}$ | $(HSH) \otimes S$ |
| 16 | $\{YX, YY, ZY\}$ | $S \otimes (SH)$ | 34 | $\{XX, XY, ZX\}$ | $(HSH) \otimes (SH)$ |
| 17 | $\{YY, YZ, ZZ\}$ | $S \otimes (HS)$ | 35 | $\{XY, XZ, YX\}$ | $(HSH) \otimes (HS)$ |
| 18 | $\{YX, YZ, ZZ\}$ | $S \otimes H$ | 36 | $\{XX, XZ, YX\}$ | $(HSH) \otimes H$ |

Table II: All 36 LU-equivalent 3-measurement sets generated from the canonical L-shaped triple $\{XX, XZ, ZX\}$ under local Cliffords $U_A \otimes U_B$.

### 4. Three unadjacent correlators, see Fig. 2 (d).

For the measurement set $\mathcal{O} = \{XX, XY, ZZ\}$, consider an observable

$$S = \alpha\,XX + \beta\,XY + \gamma\,ZZ. \tag{44}$$

Then the coefficient matrix is

$$C = \begin{pmatrix} \alpha & \beta & 0 \\ 0 & 0 & 0 \\ 0 & 0 & \gamma \end{pmatrix}, \qquad \|C\|_\infty = \max_{\alpha,\beta,\gamma} \left\{ \sqrt{\alpha^2 + \beta^2}, |\gamma| \right\}. \tag{45}$$

The expectation value is

$$\mathrm{tr}[S\rho] = \alpha \langle XX \rangle_\rho + \beta \langle XY \rangle_\rho + \gamma \langle ZZ \rangle_\rho. \tag{46}$$

Thus the normalized estimation with condition $\|C\|_\infty = 1$ is given by

$$\mathrm{N}\mathbb{E}_\rho[\mathcal{O}] = \max |\alpha\,a + \beta\,b + \gamma\,c|, \qquad a = \langle XX \rangle_\rho,\ b = \langle XY \rangle_\rho,\ c = \langle ZZ \rangle_\rho. \tag{47}$$

Upon inspection, the optimization separates into

$$\max_{\sqrt{\alpha^2 + \beta^2} = 1} |\alpha a + \beta b| = \sqrt{a^2 + b^2}, \qquad \max_{|\gamma| = 1} |\gamma c| = |c|. \tag{48}$$

Therefore we obtain the exact value

$$\mathbb{NE}_\rho[\mathcal{O}] = \sqrt{a^2 + b^2} + |c|. \tag{49}$$

Note that measurement set $\{XX, XY, YZ\}$ also leads to the same $\|C\|_\infty$ value so we can obtain the same $\mathbb{NE}_\rho[\mathcal{O}]$. The table III below shows all 3-measurement sets locally equivalent to $\{XX, XY, ZZ\}$ or $\{XX, XY, YZ\}$:

| # | Measurement set $\mathcal{O}$ | $U_A \otimes U_B$ | Measurement set $\mathcal{O}$ | $U_A \otimes U_B$ |
|---|---|---|---|---|
| 1 | $\{XX, XY, ZZ\}$ | $I \otimes I$ | $\{XX, XY, YZ\}$ | $I \otimes I$ |
| 2 | $\{XX, XY, YZ\}$ | $(HSH) \otimes I$ | $\{XX, XZ, YY\}$ | $I \otimes (HSH)$ |
| 3 | $\{XX, XZ, YY\}$ | $(HSH) \otimes (HSH)$ | $\{XY, XZ, YX\}$ | $I \otimes (SH)$ |
| 4 | $\{XX, XZ, ZY\}$ | $I \otimes (HSH)$ | $\{XX, XY, ZZ\}$ | $(HSH) \otimes I$ |
| 5 | $\{XX, YY, YZ\}$ | $(SH) \otimes (SH)$ | $\{XX, XZ, ZY\}$ | $(HSH) \otimes (HSH)$ |
| 6 | $\{XX, ZY, ZZ\}$ | $H \otimes (SH)$ | $\{XY, XZ, ZX\}$ | $(HSH) \otimes (SH)$ |
| 7 | $\{XY, XZ, YX\}$ | $(HSH) \otimes (SH)$ | $\{XZ, YX, YY\}$ | $S \otimes I$ |
| 8 | $\{XY, XZ, ZX\}$ | $I \otimes (SH)$ | $\{XY, YX, YZ\}$ | $S \otimes (HSH)$ |
| 9 | $\{XY, YX, YZ\}$ | $(SH) \otimes (HSH)$ | $\{XX, YY, YZ\}$ | $S \otimes (SH)$ |
| 10 | $\{XY, ZX, ZZ\}$ | $H \otimes (HSH)$ | $\{YX, YY, ZZ\}$ | $(SH) \otimes I$ |
| 11 | $\{XZ, YX, YY\}$ | $(SH) \otimes I$ | $\{YX, YZ, ZY\}$ | $(SH) \otimes (HSH)$ |
| 12 | $\{XZ, ZX, ZY\}$ | $H \otimes I$ | $\{YY, YZ, ZX\}$ | $(SH) \otimes (SH)$ |
| 13 | $\{YX, YY, ZZ\}$ | $S \otimes I$ | $\{XZ, ZX, ZY\}$ | $(HS) \otimes I$ |
| 14 | $\{YX, YZ, ZY\}$ | $S \otimes (HSH)$ | $\{XY, ZX, ZZ\}$ | $(HS) \otimes (HSH)$ |
| 15 | $\{YX, ZY, ZZ\}$ | $(HS) \otimes (SH)$ | $\{XX, ZY, ZZ\}$ | $(HS) \otimes (SH)$ |
| 16 | $\{YY, YZ, ZX\}$ | $S \otimes (SH)$ | $\{YZ, ZX, ZY\}$ | $H \otimes I$ |
| 17 | $\{YY, ZX, ZZ\}$ | $(HS) \otimes (HSH)$ | $\{YY, ZX, ZZ\}$ | $H \otimes (HSH)$ |
| 18 | $\{YZ, ZX, ZY\}$ | $(HS) \otimes I$ | $\{YX, ZY, ZZ\}$ | $H \otimes (SH)$ |

Table III: All 18 LU-distinct 3-measurement sets, which are locally equivalent to $\{XX, XY, ZZ\}$ on the left column, and which are locally equivalent to $\{XX, XY, YZ\}$ on the right column, via conjugation of one representative local Clifford $U_A \otimes U_B$ for each.

## II.   ENTANGLEMENT VERIFICATION FROM EXPERIMENTAL DATA

*Experimental Demonstrations.* Let us present a proof-of-principle demonstration of entanglement detection. We prepare and measure two-qubit states prepared with photon-polarization degrees of freedom. Horizontal (H) and vertical (V) polarizations are denoted as the computational basis, $|0\rangle$ and $|1\rangle$, respectively. We here demonstrate the construction of multiple EWs from the tomographically incomplete measurements.

Two types of states are considered. We have experimentally prepared states in the following, with state fidelities 98% and 95%, respectively:

- Local basis rotation about the $Y$-axis.

$$|\chi_1\rangle = (I \otimes R_Y(\theta))|\phi^+\rangle, \quad R_Y(\theta) = \exp[-i\theta Y/2].$$

- Local basis rotation about the $(1/2, 0, 1/2)$-axis.

$$|\chi_3\rangle = (I \otimes V)|\phi^+\rangle, \quad V = \frac{1}{\sqrt{2}}(\mathbb{I} + i(\cos\theta X + \sin\theta Z)).$$

Both are maximally entangled states where one of the local bases is rotated by different angles. For each state family, experiments were performed for angles ranging from $-\pi$ to $+\pi$ in increments of $\pi/9$.

### A.   Constructing entanglement witnesses

The following tables present EWs constructed from the experimental data for states where $\mathbb{NE}_\rho[\mathcal{O}] > 1$, demonstrating entanglement detection with a limited number of measurements. The significance of these results lies in the observation that for each state, the normailzed estimation $\mathbb{NE}_\rho[\mathcal{O}]$ increases when using 3 measurements compared to 2, and 4 measurements compared to 3, within the given set of 4 measurements. Furthermore, the pairs of Pauli correlations that achieve this verification are diverse, showing the versatility of the approach.

| State | EW | $\mathbb{NE}_\rho$ | State | EW | $\mathbb{NE}_\rho$ |
|---|---|---|---|---|---|
| $\left\|\chi_3(\tfrac{7}{9}\pi)\right\rangle$ | $-0.71\,XY + 0.71\,XZ + 0.71\,ZY + 0.71\,ZZ$ | 1.70 | $\left\|\chi_3(-\tfrac{7}{9}\pi)\right\rangle$ | $0.73\,YX + 0.69\,YZ - 0.69\,ZX + 0.73\,ZZ$ | 1.70 |
| | $-0.73\,XY + 0.46\,ZY + 0.73\,ZZ$ | 1.23 | | $0.74\,YX + 0.45\,YZ + 0.74\,ZZ$ | 1.25 |
| | $-XY + ZZ$ | 1.20 | | $YX + ZZ$ | 1.24 |
| $\left\|\chi_3(-\tfrac{7}{9}\pi)\right\rangle$ | $0.71\,XX + 0.70\,XY - 0.70\,ZX + 0.71\,ZY$ | 1.61 | $\left\|\chi_3(-\tfrac{2}{9}\pi)\right\rangle$ | $0.70\,XY + 0.71\,XZ - 0.71\,ZY + 0.70\,ZZ$ | 1.59 |
| | $0.74\,XX + 0.45\,XY + 0.74\,ZY$ | 1.17 | | $0.73\,XY - 0.47\,ZY + 0.73\,ZZ$ | 1.15 |
| | $XX + ZY$ | 1.15 | | $XZ - ZY$ | 1.13 |
| $\left\|\chi_3(-\tfrac{7}{9}\pi)\right\rangle$ | $-0.68\,YY + 0.74\,YZ + 0.74\,ZY + 0.68\,ZZ$ | 1.57 | $\left\|\chi_3(\tfrac{7}{9}\pi)\right\rangle$ | $0.67\,XX - 0.74\,XY - 0.74\,YX - 0.67\,YY$ | 1.54 |
| | $0.75\,YZ + 0.75\,ZY + 0.43\,ZZ$ | 1.16 | | $0.44\,XX - 0.75\,XY - 0.75\,YX$ | 1.15 |
| | $YZ + ZY$ | 1.15 | | $-XY - YX$ | 1.14 |
| $\left\|\chi_3(-\tfrac{7}{9}\pi)\right\rangle$ | $0.78\,XX - 0.63\,XZ + 0.63\,YX + 0.78\,YZ$ | 1.60 | $\left\|\chi_3(\tfrac{2}{9}\pi)\right\rangle$ | $-0.64\,YX - 0.77\,YZ - 0.77\,ZX + 0.64\,ZZ$ | 1.38 |
| | $0.75\,XX + 0.43\,YX + 0.75\,YZ$ | 1.25 | | $-0.63\,YX - 0.60\,YZ + 0.63\,ZZ$ | 1.06 |
| | $XX + YZ$ | 1.25 | | $-YZ - ZX$ | 1.06 |
| $\left\|\chi_3(\tfrac{3}{9}\pi)\right\rangle$ | $0.77\,XX - 0.64\,XY - 0.64\,ZX - 0.77\,ZY$ | 1.38 | $\left\|\chi_3(-\tfrac{6}{9}\pi)\right\rangle$ | $0.64\,XX - 0.76\,XZ + 0.76\,YX + 0.64\,YZ$ | 1.34 |
| | $0.74\,XX - 0.45\,XY - 0.74\,ZY$ | 1.08 | | $0.63\,XX + 0.61\,YX + 0.63\,YZ$ | 1.05 |
| | $XX - ZY$ | 1.06 | | $-XZ + YX$ | 1.03 |
| $\left\|\chi_3(\tfrac{2}{9}\pi)\right\rangle$ | $-0.68\,YX - 0.73\,YZ - 0.73\,ZX + 0.68\,ZZ$ | 1.33 | $\left\|\chi_1(-\tfrac{2}{9}\pi)\right\rangle$ | $0.76\,XX + 0.65\,XZ - 0.65\,YX + 0.76\,YZ$ | 1.35 |
| | $-0.65\,YX - 0.57\,YZ + 0.65\,ZZ$ | 1.04 | | $0.72\,XX + 0.48\,XZ + 0.72\,YZ$ | 1.07 |
| | $-YZ - ZX$ | 0.97 | | $XX + YZ$ | 1.03 |
| $\left\|\chi_3(-\tfrac{8}{9}\pi)\right\rangle$ | $0.73\,YX + 0.68\,YZ - 0.68\,ZX + 0.73\,ZZ$ | 1.33 | $\left\|\chi_1(\tfrac{6}{9}\pi)\right\rangle$ | $-0.71\,XX - 0.70\,XY + 0.70\,ZX - 0.71\,ZY$ | 1.27 |
| | $0.70\,YX + 0.51\,YZ + 0.70\,ZZ$ | 1.04 | | $-0.68\,XX + 0.53\,ZX - 0.68\,ZY$ | 0.99 |
| | $YX + ZZ$ | 0.97 | | $-XX - ZY$ | 0.90 |

Table IV: Entanglement witnesses constructed from experimental data for states with $\mathbb{NE}_\rho[\mathcal{O}] > 1$. Each row shows a state and its corresponding EW with different numbers of Pauli measurements (4, 3, and 2), along with the corresponding $\mathbb{NE}_\rho$ values.

| state | EW | $\mathbb{NE}_\rho$ | state | EW | $\mathbb{NE}_\rho$ |
|---|---|---|---|---|---|
| $\lvert\chi_1(-\tfrac{7}{9}\pi)\rangle$ | $-0.80\,YX + 0.60\,YZ - 0.60\,ZX - 0.80\,ZZ$ | 1.33 | $\lvert\chi_3(-\tfrac{8}{9}\pi)\rangle$ | $0.76\,XX + 0.64\,XY + 0.64\,YX - 0.76\,YY$ | 1.24 |
| | $-0.75\,YX - 0.43\,ZX - 0.75\,ZZ$ | 1.07 | | $0.62\,XX + 0.62\,XY + 0.62\,YX$ | 0.98 |
| | $-YX - ZZ$ | 1.06 | | $XX - YY$ | 0.95 |
| $\lvert\chi_3(\tfrac{7}{9}\pi)\rangle$ | $0.73\,XX - 0.68\,YX + 0.73\,ZY + 0.68\,ZZ$ | 1.95 | $\lvert\chi_3(\tfrac{7}{9}\pi)\rangle$ | $0.72\,XX - 0.69\,XY + 0.68\,YZ + 0.73\,ZZ$ | 1.88 |
| | $1.00\,XX + 0.73\,ZY + 0.68\,ZZ$ | 1.71 | | $0.72\,XX - 0.69\,XY + 1.00\,ZZ$ | 1.63 |
| | $XX + ZY$ | 1.43 | | $XX + ZZ$ | 1.36 |
| $\lvert\chi_3(\tfrac{7}{9}\pi)\rangle$ | $-0.74\,XY - 0.69\,YX + 0.72\,YZ + 0.67\,ZY$ | 1.76 | $\lvert\chi_3(-\tfrac{6}{9}\pi)\rangle$ | $-0.71\,YY + 0.70\,YZ + 0.70\,ZY + 0.71\,ZZ$ | 1.34 |
| | $-1.00\,XY - 0.69\,YX + 0.72\,YZ$ | 1.53 | | $0.64\,YZ + 0.64\,ZY + 0.59\,ZZ$ | 1.11 |
| | $-XY + YZ$ | 1.29 | | $-YY + ZZ$ | 0.96 |
| $\lvert\chi_1(\tfrac{2}{9}\pi)\rangle$ | $0.58\,XX - 0.81\,XY + 0.81\,ZX + 0.58\,ZY$ | 1.29 | $\lvert\chi_3(\tfrac{6}{9}\pi)\rangle$ | $0.74\,XX - 0.67\,XY + 0.67\,ZX + 0.74\,ZY$ | 1.24 |
| | $0.45\,XX - 0.74\,XY + 0.74\,ZX$ | 1.07 | | $0.68\,XX - 0.54\,XY + 0.68\,ZY$ | 1.02 |
| | $-XY + ZX$ | 1.05 | | $XX + ZY$ | 0.92 |
| $\lvert\chi_3(-\tfrac{7}{9}\pi)\rangle$ | $0.69\,XX + 0.72\,YX + 0.66\,ZY + 0.75\,ZZ$ | 1.88 | $\lvert\chi_3(\tfrac{7}{9}\pi)\rangle$ | $0.76\,XX - 0.65\,YX + 0.67\,ZY + 0.74\,ZZ$ | 1.82 |
| | $0.69\,XX + 0.72\,YX + 1.00\,ZZ$ | 1.66 | | $1.00\,XX + 0.67\,ZY + 0.74\,ZZ$ | 1.60 |
| | $YX + ZZ$ | 1.38 | | $XX + ZZ$ | 1.36 |
| $\lvert\chi_3(-\tfrac{7}{9}\pi)\rangle$ | $0.67\,XX + 0.75\,YX + 0.70\,ZY + 0.72\,ZZ$ | 1.69 | $\lvert\chi_1(\tfrac{2}{9}\pi)\rangle$ | $-0.63\,YX + 0.77\,YZ + 0.77\,ZX + 0.63\,ZZ$ | 1.28 |
| | $1.00\,YX + 0.70\,ZY + 0.72\,ZZ$ | 1.47 | | $0.70\,YZ + 0.70\,ZX + 0.51\,ZZ$ | 1.06 |
| | $YX + ZZ$ | 1.24 | | $YZ + ZX$ | 0.99 |
| $\lvert\chi_3(-\tfrac{7}{9}\pi)\rangle$ | $0.77\,XY + 0.65\,YX + 0.76\,YZ + 0.64\,ZY$ | 1.89 | $\lvert\chi_3(-\tfrac{7}{9}\pi)\rangle$ | $0.70\,XX + 0.71\,XY + 0.65\,YZ + 0.76\,ZZ$ | 1.85 |
| | $1.00\,XY + 0.65\,YX + 0.76\,YZ$ | 1.68 | | $0.70\,XX + 0.71\,XY + 1.00\,ZZ$ | 1.64 |
| | $XY + YZ$ | 1.44 | | $XY + ZZ$ | 1.35 |
| $\lvert\chi_3(\tfrac{3}{9}\pi)\rangle$ | $-0.75\,XY - 0.78\,YX - 0.62\,YZ - 0.67\,ZY$ | 1.81 | $\lvert\chi_3(-\tfrac{1}{9}\pi)\rangle$ | $0.76\,XX + 0.65\,XY + 0.65\,YX - 0.76\,YY$ | 1.17 |
| | $-0.75\,XY - 1.00\,YX - 0.67\,ZY$ | 1.61 | | $0.63\,XX + 0.61\,XY + 0.61\,YX$ | 0.97 |
| | $-XY - YX$ | 1.38 | | $XX - YY$ | 0.88 |
| $\lvert\chi_3(-\tfrac{7}{9}\pi)\rangle$ | $0.63\,XX + 0.77\,YX + 0.78\,YZ + 0.63\,ZX$ | 1.86 | $\lvert\chi_1(-\tfrac{3}{9}\pi)\rangle$ | $0.69\,XX + 0.73\,XY - 0.73\,ZX + 0.69\,ZY$ | 1.19 |
| | $1.00\,XY + 0.78\,YZ + 0.63\,ZZ$ | 1.65 | | $0.63\,XX - 0.60\,ZX + 0.63\,ZY$ | 0.99 |
| | $XY + YZ$ | 1.44 | | $XX - ZX$ | 0.86 |
| $\lvert\chi_3(\tfrac{6}{9}\pi)\rangle$ | $0.68\,XX + 0.73\,XZ - 0.73\,YX + 0.68\,YZ$ | 1.21 | $\lvert\chi_3(-\tfrac{7}{9}\pi)\rangle$ | $0.76\,XY + 0.78\,YX + 0.62\,YZ + 0.65\,ZY$ | 1.82 |
| | $0.62\,XX - 0.61\,YX + 0.62\,YZ$ | 1.01 | | $0.76\,XY + 1.00\,YX + 0.65\,ZY$ | 1.62 |
| | $-XZ - YX$ | 0.89 | | $XY + YX$ | 1.40 |
| $\lvert\chi_3(-\tfrac{3}{9}\pi)\rangle$ | $-0.69\,YY - 0.72\,YZ - 0.72\,ZY + 0.69\,ZZ$ | 1.19 | $\lvert\chi_3(\tfrac{3}{9}\pi)\rangle$ | $-0.62\,XY - 0.81\,YX - 0.59\,YZ - 0.78\,ZY$ | 1.94 |
| | $-0.65\,YZ - 0.65\,ZY + 0.58\,ZZ$ | 0.99 | | $-0.62\,XY - 1.00\,YX - 0.78\,ZY$ | 1.75 |
| | $-YZ - ZY$ | 0.85 | | $-YX - ZY$ | 1.54 |
| $\lvert\chi_1(\tfrac{1}{9}\pi)\rangle$ | $-0.81\,YX + 0.59\,YZ + 0.59\,ZX + 0.81\,ZZ$ | 1.21 | $\lvert\chi_3(\tfrac{5}{9}\pi)\rangle$ | $-0.79\,YX - 0.62\,YY + 0.62\,ZX - 0.79\,ZY$ | 1.18 |
| | $0.55\,YZ + 0.55\,ZX + 0.70\,ZZ$ | 1.02 | | $-0.67\,YX - 0.57\,YY + 0.57\,ZX$ | 0.99 |
| | $-YX + ZZ$ | 0.97 | | $-YX - ZY$ | 0.93 |
| $\lvert\chi_1(-\tfrac{2}{9}\pi)\rangle$ | $-0.61\,XY + 0.79\,XZ + 0.79\,ZY + 0.61\,ZZ$ | 1.25 | $\lvert\chi_1(\tfrac{6}{9}\pi)\rangle$ | $-0.72\,XX - 0.69\,XZ - 0.69\,YX + 0.72\,YZ$ | 1.13 |
| | $0.70\,XZ + 0.70\,ZY + 0.51\,ZZ$ | 1.07 | | $-0.65\,XX - 0.58\,XZ + 0.65\,YZ$ | 0.95 |
| | $XZ + ZY$ | 1.00 | | $-XX + YZ$ | 0.82 |
| $\lvert\chi_3(-\tfrac{2}{9}\pi)\rangle$ | $0.67\,XY + 0.58\,YX - 0.82\,YZ - 0.74\,ZY$ | 1.98 | $\lvert\chi_1(-\tfrac{7}{9}\pi)\rangle$ | $-0.66\,XX + 0.76\,XY - 0.76\,ZX - 0.66\,ZY$ | 1.16 |
| | $0.67\,XY - 1.00\,YZ - 0.74\,ZY$ | 1.79 | | $-0.55\,XX + 0.67\,XY - 0.67\,ZX$ | 0.99 |
| | $-YZ - ZY$ | 1.54 | | $XY - ZX$ | 0.88 |
| $\lvert\chi_3(\tfrac{2}{9}\pi)\rangle$ | $0.64\,XX - 0.77\,XY - 0.77\,YX - 0.64\,YY$ | 1.16 | $\lvert\chi_3(\tfrac{3}{9}\pi)\rangle$ | $0.62\,XX - 0.78\,XY - 0.65\,YZ + 0.76\,ZZ$ | 1.66 |
| | $0.54\,XX - 0.68\,XY - 0.68\,YX$ | 0.99 | | $-1.00\,XY - 0.65\,YZ + 0.76\,ZZ$ | 1.49 |
| | $-XY - YX$ | 0.90 | | $-XY + ZZ$ | 1.28 |
| $\lvert\chi_3(\tfrac{3}{9}\pi)\rangle$ | $0.60\,XX - 0.80\,XY - 0.70\,YZ + 0.72\,ZZ$ | 1.65 | $\lvert\chi_3(-\tfrac{2}{9}\pi)\rangle$ | $0.80\,XY + 0.58\,YX - 0.81\,YZ - 0.60\,ZY$ | 1.83 |
| | $-1.00\,XY - 0.70\,YZ + 0.72\,ZZ$ | 1.49 | | $0.80\,XY - 1.00\,YZ - 0.60\,ZY$ | 1.66 |
| | $-XY + ZZ$ | 1.25 | | $XY - YZ$ | 1.47 |
| $\lvert\chi_3(\tfrac{7}{9}\pi)\rangle$ | $0.80\,XX - 0.60\,XY + 0.61\,YZ + 0.79\,ZZ$ | 1.72 | $\lvert\chi_3(-\tfrac{3}{9}\pi)\rangle$ | $-0.50\,YY - 0.87\,YZ - 0.87\,ZY + 0.50\,ZZ$ | 1.35 |
| | $1.00\,XX + 0.61\,YZ + 0.79\,ZZ$ | 1.56 | | $-0.74\,YZ - 0.74\,ZY + 0.45\,ZZ$ | 1.19 |
| | $XX + ZZ$ | 1.38 | | $-YZ - ZY$ | 1.17 |
| $\lvert\chi_3(-\tfrac{3}{9}\pi)\rangle$ | $0.84\,XY + 0.77\,YX - 0.64\,YZ - 0.55\,ZZ$ | 1.97 | $\lvert\chi_3(\tfrac{8}{9}\pi)\rangle$ | $-0.74\,YX + 0.67\,YZ + 0.67\,ZX + 0.74\,ZZ$ | 1.18 |
| | $1.00\,XY + 0.77\,YX - 0.64\,YZ$ | 1.81 | | $-0.65\,YX + 0.58\,YZ + 0.65\,ZZ$ | 1.03 |
| | $XY + YX$ | 1.58 | | $-YX + ZZ$ | 0.88 |

Table V: Continuation of entanglement witnesses constructed from experimental data, showing the diversity of Pauli correlation pairs that enable entanglement verification.

## B. Tomography and Pauli Correlation of Data

This subsection presents the Pauli correlation data and tomographically reconstructed density matrices from the experimental measurements via maximum-likelihood method. We show heatmaps of Pauli correlators $\langle A_i \otimes B_j \rangle_{est}$ as functions of the state parameter $\theta$, raw correlator matrices, and the reconstructed density matrices.

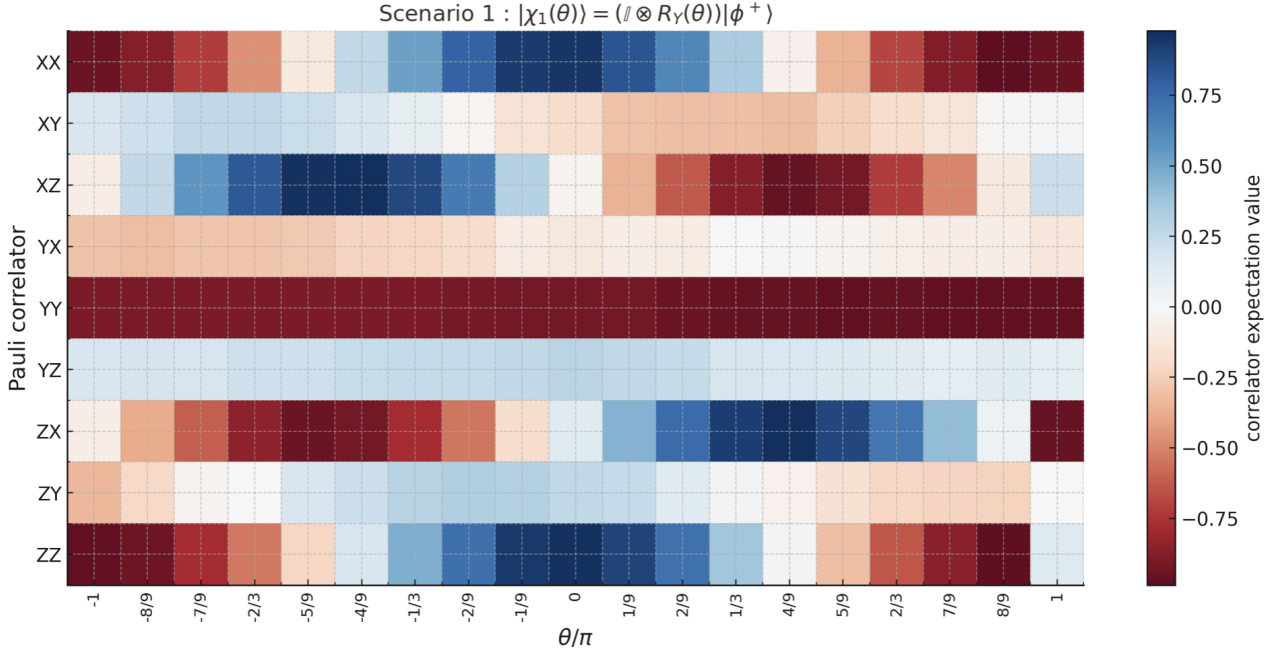

Figure 3: Heat-map of Pauli correlators $\langle A_i \otimes B_j \rangle_{est}$ for state family 1, plotted as a function of the state-family parameter $\theta$. The $x$-axis labels indicate the discrete angles in the form $(\pm k/9)\pi$ ranging from $-\pi$ to $+\pi$ in increments of $\pi/9$, while the $y$-axis shows the corresponding Pauli correlator. State family 1 is described by Eq. (II) with maximum fidelity of 98%.

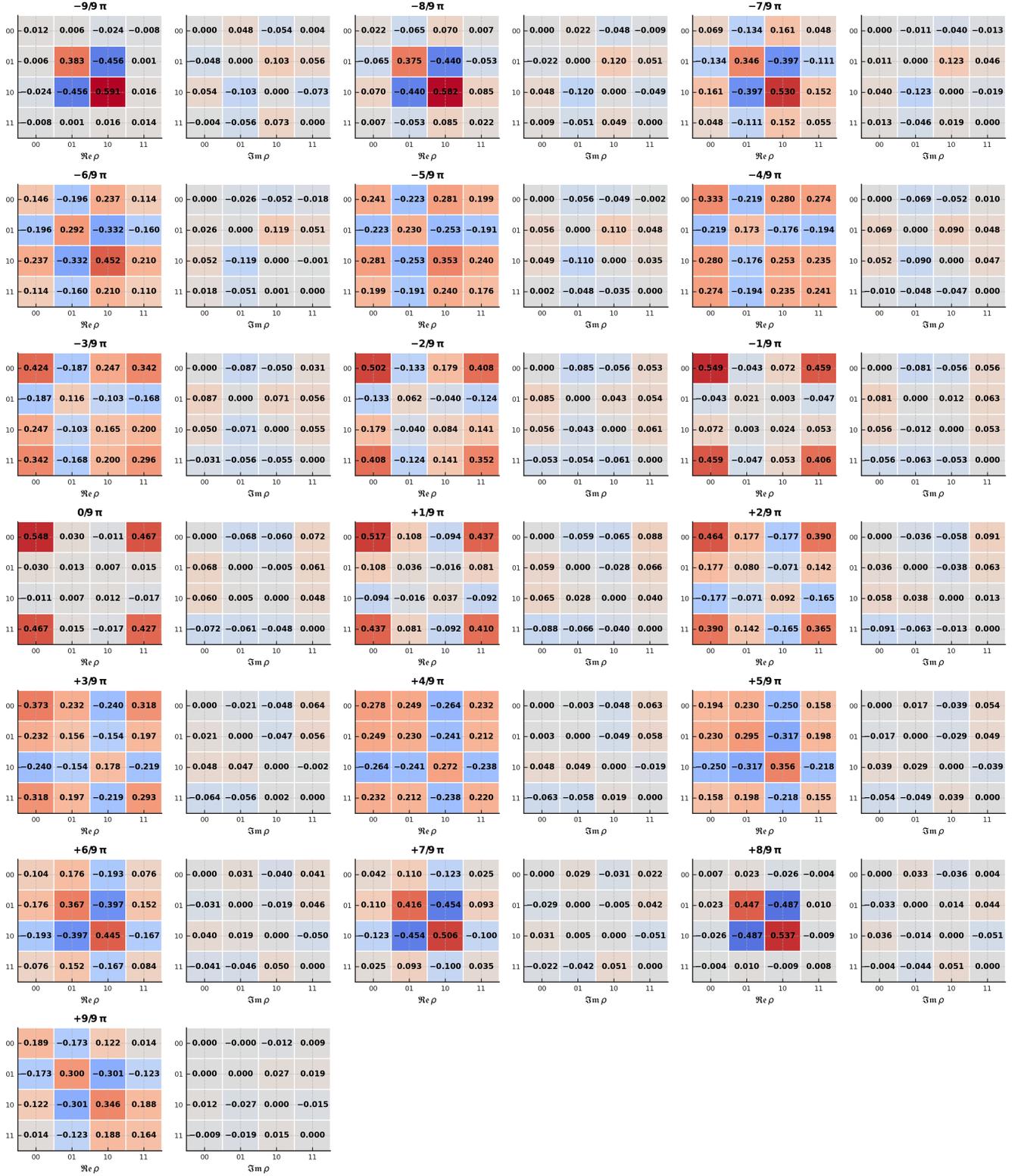

Figure 5: Maximum-likelihood–reconstructed density matrices for all angles $\theta$ in state family 1, displayed in a $4 \times 4$ panel structure for each angle. State family 1 is described by Eq. (II) with maximum fidelity of 98%, representing maximally entangled states rotated about the $Y$-axis. Angles range from $-\pi$ to $+\pi$ in increments of $\pi/9$. For every block, the **1st and 3rd columns** show the **real part** of $\rho$, $\Re\mathfrak{e}\,\rho$, while the **2nd and 4th columns** display the **imaginary part**, $\Im\mathfrak{m}\,\rho$. All heatmaps are rendered using a single global color scale shared across all angles and all panels. Titles indicate each corresponding angle in the compact form $(\pm k/9)\pi$.

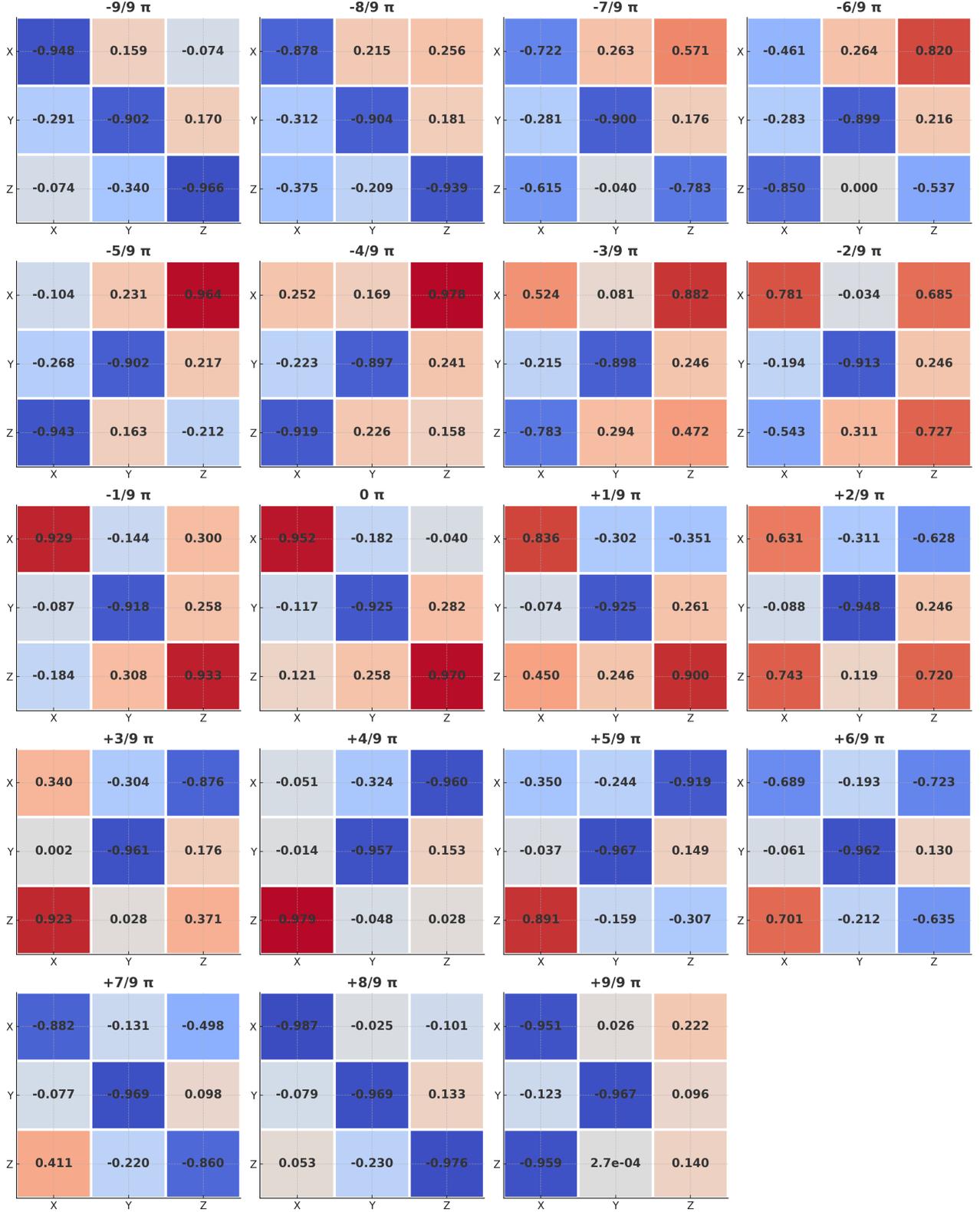

Figure 4: Heat-map style visualization of the raw data Pauli correlator matrices $\langle A_i \otimes B_j \rangle_{est}$ for state family 1 (experiment performed on April 3rd) for all angles $\theta$. Each panel uses a unified color scale and shows the correlators for $A_i, B_j \in \{X, Y, Z\}$, with values labeled in bold. Angles are ordered from $-9\pi/9$ to $+9\pi/9$ in increments of $\pi/9$. All panels share a common color normalization determined by the global maximum $|\langle A_i \otimes B_j \rangle_{est}|$ across the entire dataset.

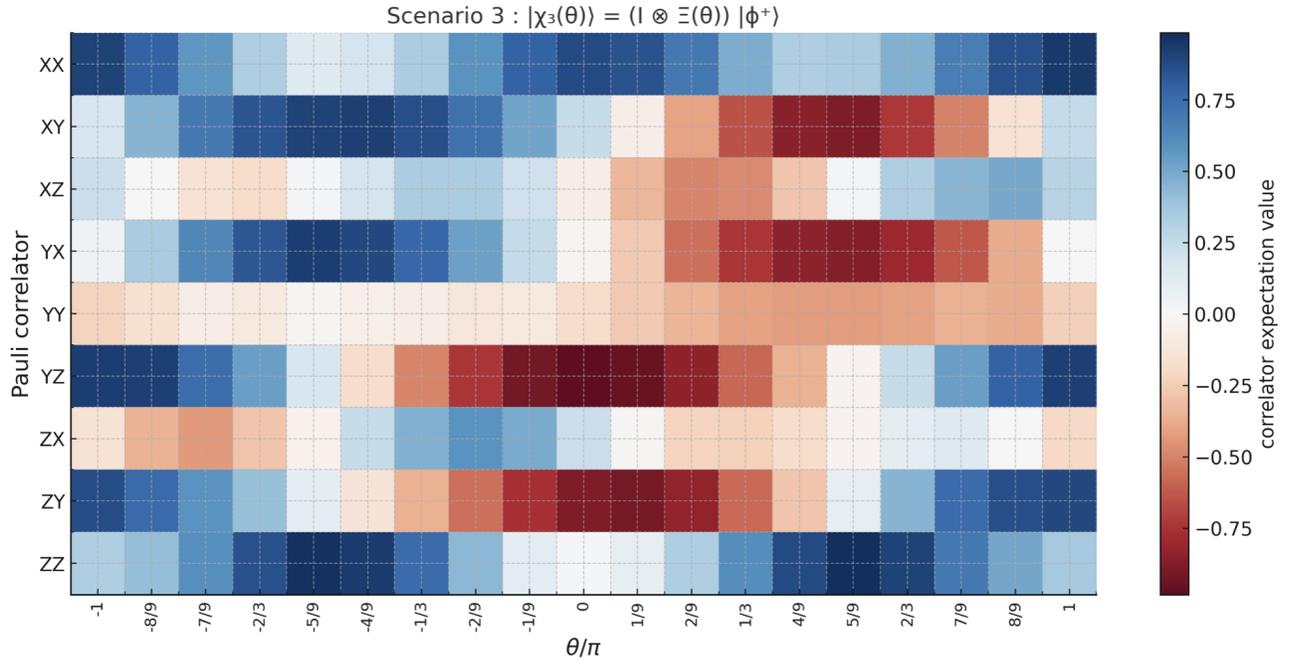

Figure 6: Heat-map of Pauli correlators $\langle A_i \otimes B_j \rangle_{est}$ for state family 3, plotted as a function of the state-family parameter $\theta$. The $x$-axis labels indicate the discrete angles in the form $(\pm k/9)\pi$ ranging from $-\pi$ to $+\pi$ in increments of $\pi/9$, while the $y$-axis shows the corresponding Pauli correlator. State family 3 is described by Eq. (II) with maximum fidelity of 95%.

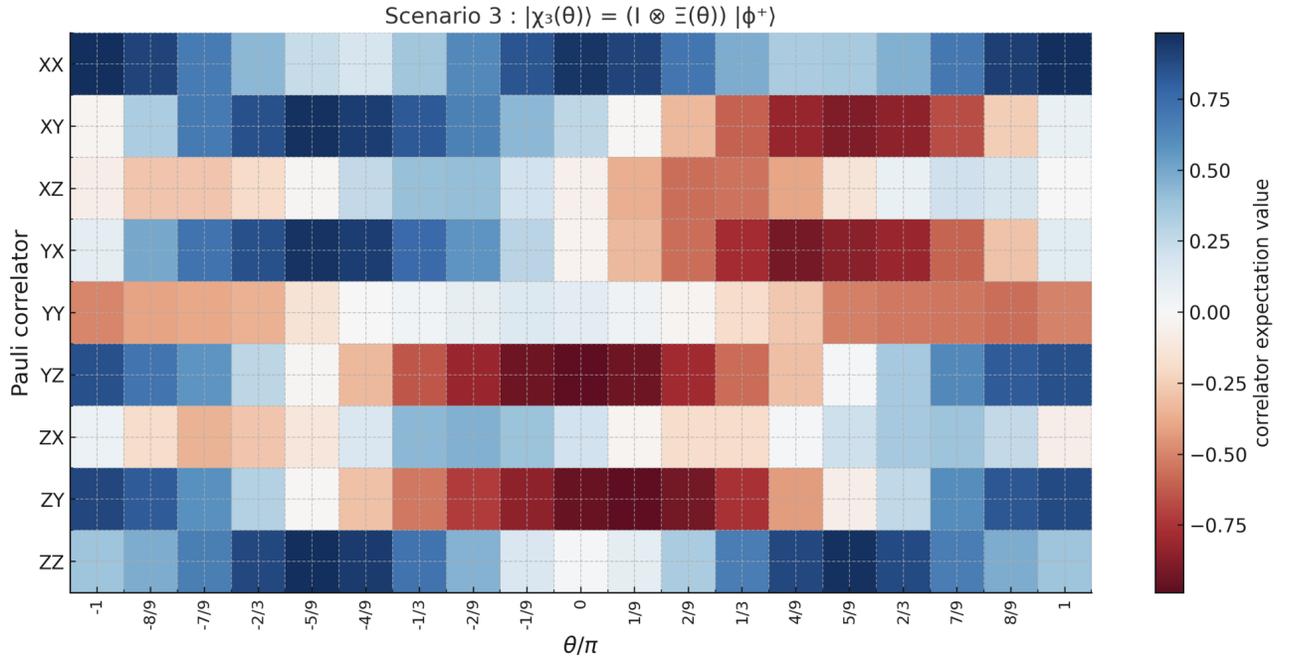

Figure 7: Heat-map of Pauli correlators $\langle A_i \otimes B_j \rangle_{est}$ for state family 3 (experiment performed on May 3rd), plotted as a function of the state-family parameter $\theta$. The $x$-axis labels indicate the discrete angles in the form $(\pm k/9)\pi$ ranging from $-\pi$ to $+\pi$ in increments of $\pi/9$, while the $y$-axis shows the corresponding Pauli correlator. State family 3 is described by Eq. (II) with maximum fidelity of 95%.

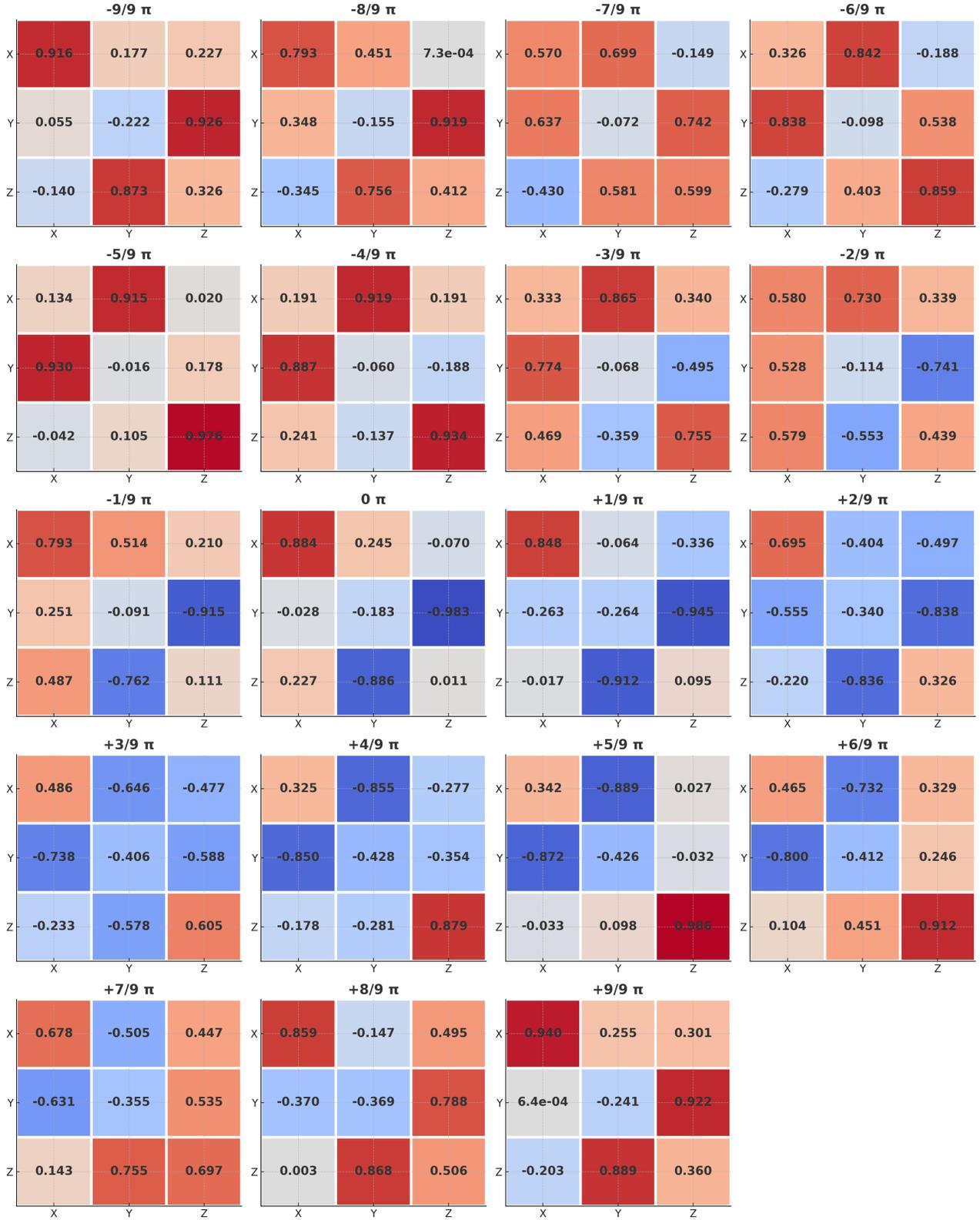

Figure 8: Heat-map style visualization of the raw data Pauli correlator matrices $\langle A_i \otimes B_j \rangle_{est}$ for state family 3 for all angles $\theta$. Each panel uses a unified color scale and shows the correlators for $A_i, B_j \in \{X, Y, Z\}$, with values labeled in bold. Angles are ordered from $-9\pi/9$ to $+9\pi/9$ in increments of $\pi/9$. All panels share a common color normalization determined by the global maximum $|\langle A_i \otimes B_j \rangle_{est}|$ across the entire dataset.

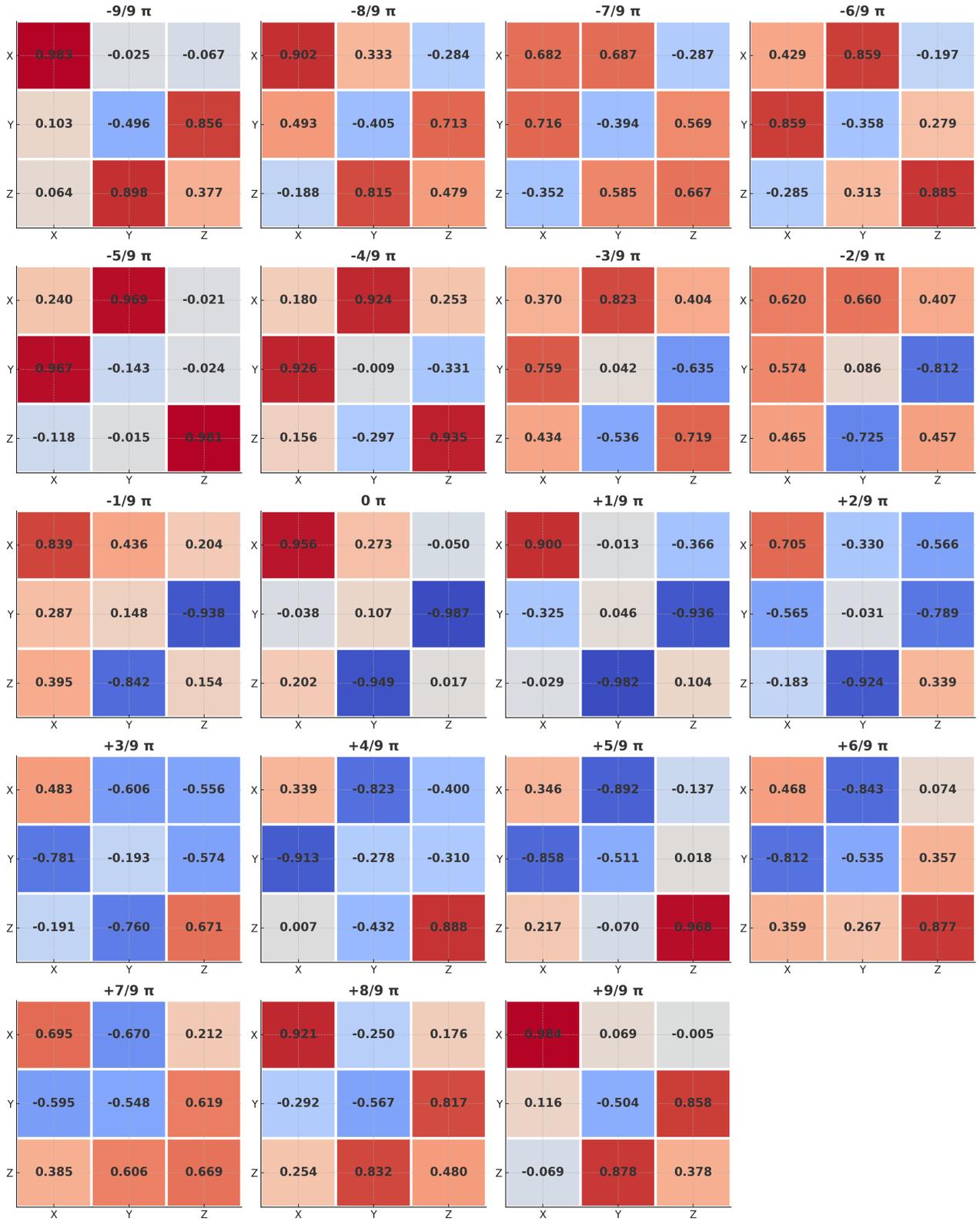

Figure 9: Heat-map style visualization of the raw data Pauli correlator matrices $\langle A_i \otimes B_j \rangle_{est}$ for state family 3 for all angles $\theta$. Each panel uses a unified color scale and shows the correlators for $A_i, B_j \in \{X, Y, Z\}$, with values labeled in bold. Angles are ordered from $-9\pi/9$ to $+9\pi/9$ in increments of $\pi/9$. All panels share a common color normalization determined by the global maximum $|\langle A_i \otimes B_j \rangle_{est}|$ across the entire dataset.

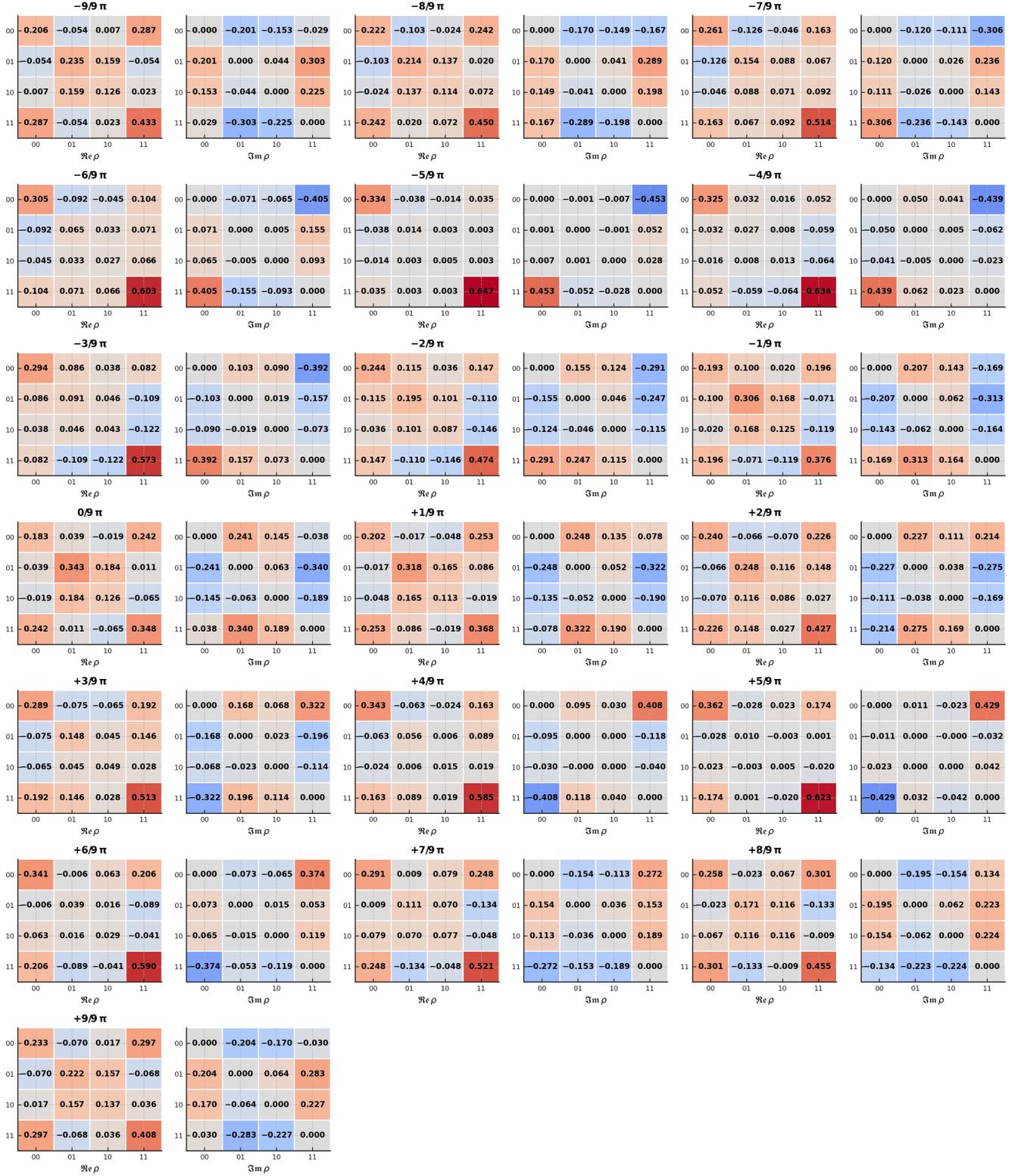

Figure 10: Maximum-likelihood–reconstructed density matrices for all angles $\theta$ in state family 3, displayed in a $4 \times 4$ panel structure for each angle. State family 3 is described by Eq. (II) with maximum fidelity of 95%, representing maximally entangled states rotated about an arbitrary axis. Angles range from $-\pi$ to $+\pi$ in increments of $\pi/9$. For every block, the **1st and 3rd columns** show the **real part** of $\rho$, $\mathfrak{Re}\,\rho$, while the **2nd and 4th columns** display the **imaginary part**, $\mathfrak{Im}\,\rho$. All heatmaps are rendered using a single global color scale shared across all angles and all panels, enabling direct visual comparison of amplitude variations. Titles indicate each corresponding angle in the compact form $(\pm k/9)\pi$.

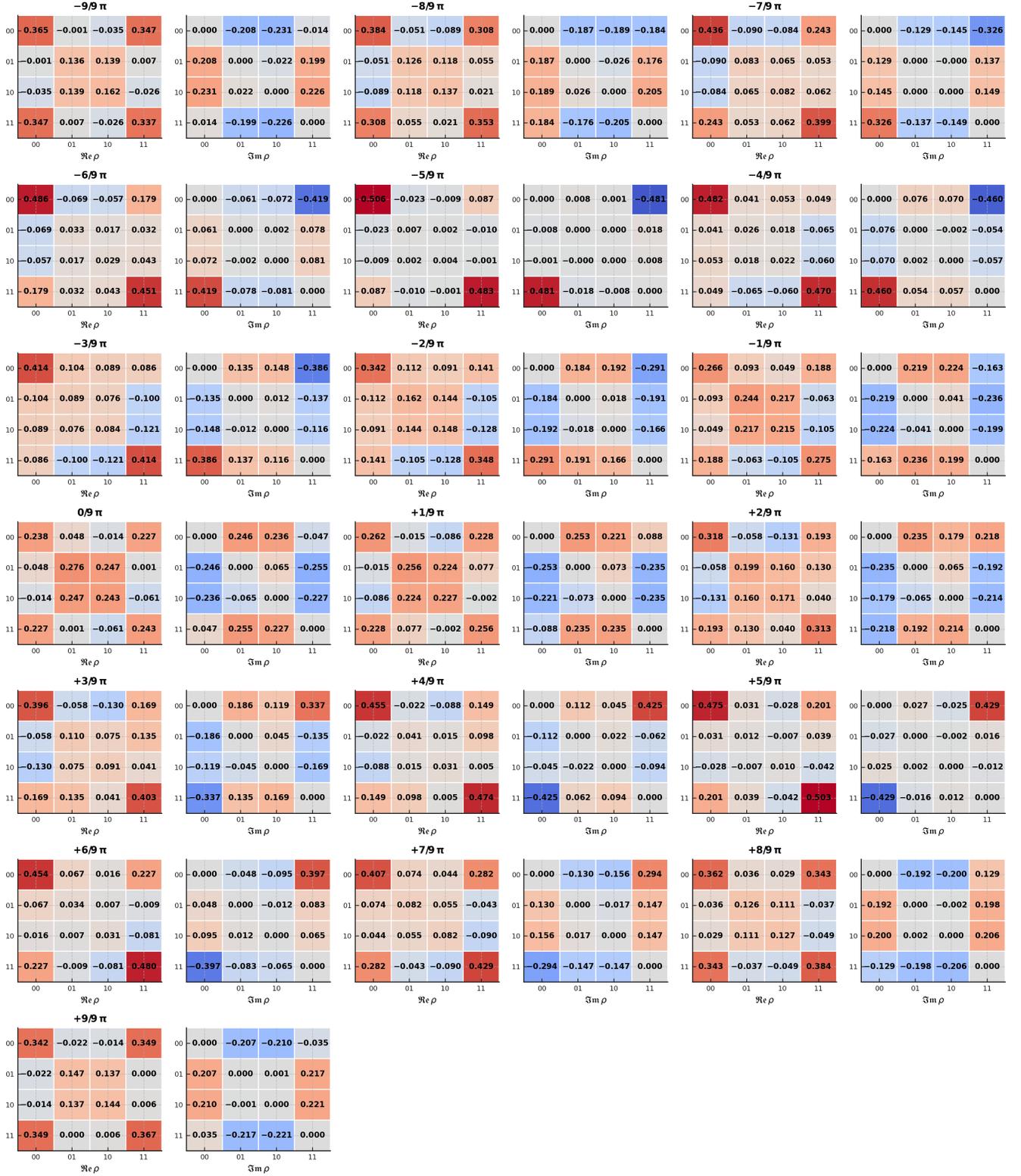

Figure 11: Maximum-likelihood–reconstructed density matrices for all angles $\theta$ in state family 3 displayed in a $4 \times 4$ panel structure for each angle. State family 3 is described by Eq. (II) with maximum fidelity of 95%, representing maximally entangled states rotated about an arbitrary axis. Angles range from $-\pi$ to $+\pi$ in increments of $\pi/9$. For every block, the **1st and 3rd columns** show the **real part** of $\rho$, $\Re\mathfrak{e}\,\rho$, while the **2nd and 4th columns** display the **imaginary part**, $\Im\mathfrak{m}\,\rho$. All heatmaps are rendered using a single global color scale shared across all angles and all panels, enabling direct visual comparison of amplitude variations. Titles indicate each corresponding angle in the compact form $(\pm k/9)\pi$.